\documentclass[lettersize,journal]{IEEEtran}
\usepackage{amsmath,amsfonts}
\usepackage{algorithmic}
\usepackage{algorithm}
\usepackage{array}
\usepackage[caption=false,font=normalsize,labelfont=sf,textfont=sf]{subfig}
\usepackage{textcomp}
\usepackage{stfloats}
\usepackage{url}
\usepackage{verbatim}
\usepackage{cite}
\usepackage[table]{xcolor}         % colors
\usepackage{xcolor}         % colors
\usepackage{hyperref}

\usepackage{todonotes}
\usepackage{url}
\usepackage{amssymb}
\usepackage{booktabs}
\usepackage{multirow}
\usepackage{caption}
\usepackage{float}
\usepackage{subcaption}
\usepackage{graphicx}
\usepackage{arydshln}

\begin{document}

\title{Hierarchical Control of Emotion Rendering in Speech Synthesis
\thanks{$\dag$ Corresponding author}
% \thanks{The research is supported by National Natural Science Foundation of China (Grant No. 62271432); Internal Project of Shenzhen Research Institute of Big Data (Grant No. T00120220002); Shenzhen Science and Technology Program ZDSYS20230626091302006; Shenzhen Science and Technology Research Fund (Fundamental Research Key Project Grant No. JCYJ20220818103001002); and CCF-NetEase ThunderFire Innovation Research Funding (No. CCF-Netease 202302).}
\thanks{
%The research is supported by National Natural Science Foundation of China (Grant No. 62271432); Shenzhen Science and Technology Program ZDSYS20230626091302006; Open Project of the Key Laboratory of Artificial Intelligence, Ministry of Education (No. AI202405); and CCF-NetEase ThunderFire Innovation Research Funding (No. CCF-Netease 202302).
The research is supported by
National Natural Science Foundation of China (Grant No. 62271432);
Shenzhen Science and Technology Program (Shenzhen Key Laboratory, Grant No. ZDSYS20230626091302006);
Program for Guangdong Introducing Innovative and Entrepreneurial Teams, Grant No. 2023ZT10X044;
Open Project of the Key Laboratory of Artificial Intelligence, Ministry of Education (No. AI202405); and CCF-NetEase ThunderFire Innovation Research Funding (No. CCF-Netease 202302).
}
}

%\author{IEEE Publication Technology,~\IEEEmembership{Staff,~IEEE,}
%        % <-this % stops a space
%\thanks{This paper was produced by the IEEE Publication Technology Group. They are in Piscataway, NJ.}% <-this % stops a space
%\thanks{Manuscript received April 19, 2021; revised August 16, 2021.}}

\author{
\IEEEauthorblockN{
Sho Inoue\textsuperscript{1,3},
Kun Zhou\textsuperscript{4},
Shuai Wang\textsuperscript{2,3$\dag$}, and
Haizhou Li\textsuperscript{1,3,5$\dag$} \\
}
\IEEEauthorblockA{
\textsuperscript{1}\textit{School of Data Science, The Chinese University of Hong Kong, Shenzhen (CUHK-Shenzhen), China} \\
%\textsuperscript{2}\textit{Key Laboratory of Artificial Intelligence, Ministry of Education, Shanghai, China}\\
\textsuperscript{2}\textit{School of Intelligence Science and Technology, Nanjing University, Suzhou, China}\\
\textsuperscript{3}\textit{Shenzhen Research Institute of Big Data, Shenzhen, China}\\
\textsuperscript{4}\textit{Tongyi Speech Lab, Alibaba Group, Singapore}\\
\textsuperscript{5}\textit{Department of ECE, National University of Singapore, Singapore}\\
shoinoue@link.cuhk.edu.cn, kun.z@alibaba-inc.com, wangshuai@cuhk.edu.cn, haizhouli@cuhk.edu.cn
}
}

\markboth{Journal of \LaTeX\ Class Files,~Vol.~14, No.~8, August~2021}%
{Shell \MakeLowercase{\textit{et al.}}: A Sample Article Using IEEEtran.cls for IEEE Journals}

\IEEEpubid{0000--0000/00\$00.00~\copyright~2021 IEEE}

\maketitle
\begin{abstract}

Emotional text-to-speech synthesis (TTS) aims to generate realistic emotional speech from input text. However, quantitatively controlling multi-level emotion rendering remains challenging. In this paper, we propose a flow-matching based emotional TTS framework with a novel approach for emotion intensity modeling to facilitate fine-grained control over emotion rendering at the phoneme, word, and utterance levels. We introduce a hierarchical emotion distribution (ED) extractor that captures a quantifiable ED embedding across different speech segment levels. Additionally, we explore various acoustic features and assess their impact on emotion intensity modeling. During TTS training, the hierarchical ED embedding effectively captures the variance in emotion intensity from the reference audio and correlates it with linguistic and speaker information. The TTS model not only generates emotional speech during inference, but also quantitatively controls the emotion rendering over the speech constituents. Both objective and subjective evaluations demonstrate the effectiveness of our framework in terms of speech quality, emotional expressiveness, and hierarchical emotion control.

\end{abstract}


\section{Introduction} \label{sec:introduction}

%\fi

% Here, the introduction of emotional tts system. And, mention that the multi-level is still under-explored
%\textcolor{red}{Hi Sho, can you read the abstract and introduction to see if there is any misunderstanding from my side? If you are happy with it, can you also fill the citations?}
\IEEEPARstart{N}eural text-to-speech (TTS) systems have significantly improved the naturalness of synthesized speech but still struggle with human-like emotional expressions~\cite{triantafyllopoulos2023overview,schuller2018age}. We focus on emotional TTS research to produce realistic emotional speech from text input. Emotional TTS has become a vital technology in human-computer interactions~\cite{pittermann2010handling,kun2022emotion}, providing more responsive and natural communication. Emotional TTS supports diverse applications, including virtual assistants, customer support systems, and interactive gaming~\cite{Triantafyllopoulos2022AnOO}.

The major challenge of emotional TTS is the modeling of the one-to-many function. Unlike traditional TTS systems, which primarily focus on capturing phonetic variations across speakers~\cite{fastspeech2,zhou24_interspeech,rltts}, emotional TTS aims to render diverse emotional styles over text~\cite{schroder2001emotional}. As a form of speech expressiveness, speech emotion not only correlates to multiple prosodic features including pitch, energy and speech rate~\cite{tan2021survey,hirschberg2004pragmatics,belyk2014perception}, but also varies among individuals and languages~\cite{Laukka2017,zhou2021vaw,zhou2020converting}. Emotional TTS studies have benefited from advances in emotional representation learning, evolving from hand-crafted features in tools like OpenSMILE~\cite{opensmile} to deep emotional features~\cite{ma2023emotion2vec} and self-supervised learning (SSL) representations~\cite{WavLM,Hubert}. The improved generalizability of these features enhances the performance of emotional TTS systems. However, controlling the emotion rendering remains a challenging topic. 
%Previous studies leverage pre-defined emotional features set, for example, OpenSMILE, to model emotions. later studies 
\begin{figure}[t]
  \centering
  \centerline{\includegraphics[width=8.2cm]{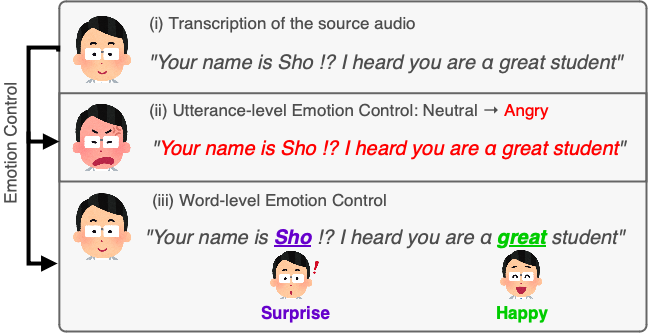}}
  %\vspace{-0.3cm}
 \caption{
An example of hierarchical emotion rendering control shows that emotions can be applied either across the entire utterance (ii) or vary at the level of individual words (iii).
    }
 \label{fig:intro}
\end{figure}
\IEEEpubidadjcol

Previous emotional TTS frameworks typically treat emotion as a global speech attribute~\cite{emogst,wu2019endtoend}.
However, these approaches often lead to synthesized voices with average emotional styles, providing limited control over emotion. Few studies have focused on controlling emotion intensities by manipulating emotion embedding through interpolation or scaling~\cite{emoscaler,oh2023semisupervised,zhang2023iemotts,li2022crossspeaker}. Recent research proposes to model emotion as a relative attribute~\cite{parikh2011relative} by comparing neutral and emotional speech pairs~\cite{kun-intensity,9003829,kun-mix,msemotts,ShoICASSP,ShoArxiv}. This method allows for more precise control of emotion intensity~\cite{kun-intensity,9003829} and facilitates the creation of mixed emotional styles~\cite{kun-mix}. However, the above studies are designed to handle emotions primarily at the utterance level, which does not sufficiently accommodate the finer nuances of emotion that can vary significantly within a single utterance due to context, phrasing, or individual words. The hierarchical structure of speech emotions, characterized by prosodic patterns at the phoneme, word, and utterance levels~\cite{hieprosody,zhou2020transforming}, includes global prosodic elements such as tempo and speaker traits~\cite{hy-utt}, as well as local factors like pitch and lexical emphasis~\cite{hy-word1,hy-word2,hy-word3,hy-ph1,hy-ph2,hy-ph3}. These elements are crucial for conveying the subtleties of emotional nuances. Therefore, a quantitative approach to controlling multi-level emotion rendering is essential.

In this paper, motivated by prior studies that emphasize the hierarchical nature of speech emotions \cite{hieprosody,hy-utt,hy-word1}, we introduce a novel emotional TTS framework that explicitly adopts hierarchical emotion modeling and futher enhance the run-time emotion control. We enable users to flexibly adjust not only the style and intensity of emotions, but also their modulation at the phoneme, phrase, or sentence level. Our approach focuses on capturing and controlling the complex emotional layers and transitions that naturally occur in human speech. During training, the framework leverages reference audio to learn hierarchical emotional variations and correlate them with linguistic and speaker information, allowing a fine-grained and quantifiable control over emotion rendering during inference. 
\textcolor{black}{
Compared to our previous work \cite{ShoICASSP,ShoArxiv}, we present a robust hierarchical emotion distribution modeling approach that enables finer-grained emotion representations. By integrating multi-level acoustic features, our model captures local emotional nuances through lower-level features and broader emotional contexts via higher-level self-supervised representations. Additionally, we explore the use of deep neural networks as a replacement for traditional SVM-based models, leading to improved precision and flexibility in emotion intensity modeling.
}
Our novel contributions are summarized as follows:

{
\setlength{\leftmargini}{10pt} 
\begin{itemize}
\item We introduce a flow-matching based emotional TTS framework that facilitates fine-grained and quantitative control over emotion rendering at the phoneme, word, and utterance levels;
\item We design a hierarchical emotion distribution (ED) extraction module, which captures quantifiable ED embeddings across different speech segments. This ED embedding is used to guide the TTS system to recover the reference emotion during training is used to adjust the emotion variance across the speech constituents;

\item Through comprehensive experiments, we explore different acoustic representations and emotion intensity modeling methods, evaluating their effectiveness in emotion intensity modeling and their impact on emotion control.
\end{itemize}
}

The rest of this paper is organized as follows: In Section 2, we introduce related works. Section 3 describes our proposed methodology. In Section 4, we introduce our experiment setup. In section 5, we analyze the results and Section 6 concludes our study. We placed speech demos on the project page\footnote{\textbf{Project Page}: \url{https://github.com/shinshoji01/HED-project-page}}.

\begin{figure*}[t]
  \centering
  \centerline{\includegraphics[width=17cm]{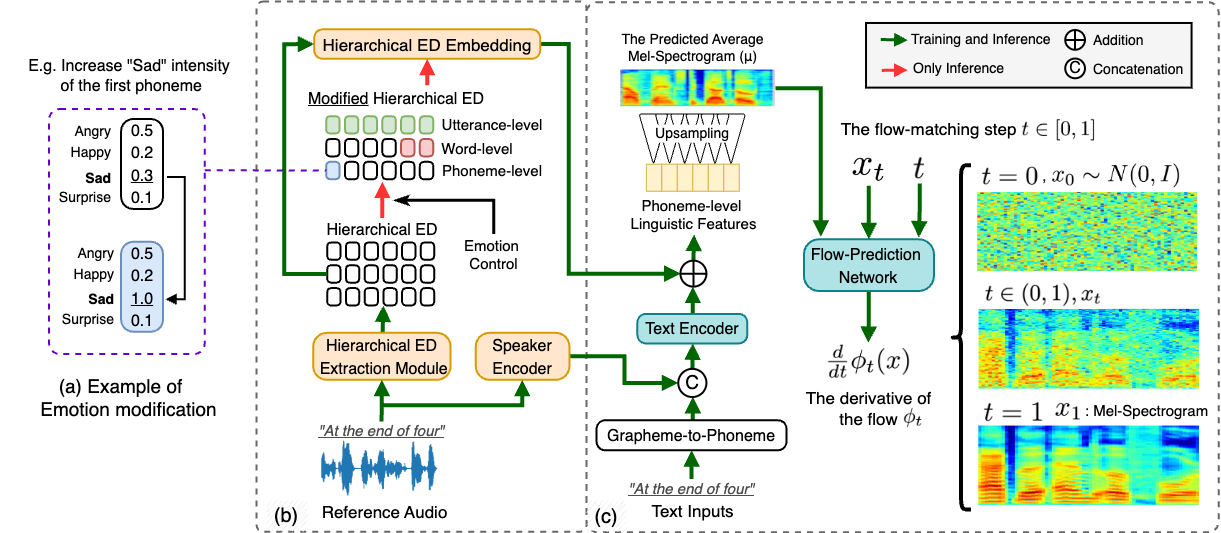}}
  %\missingfigure[figwidth=6cm]{Testing a long text string}
 \caption{The overall diagram of the proposed TTS framework. (a) An example of emotion modification (b) The style conditioning of the TTS framework including hierarchical emotion distribution (ED) and speaker embedding. (c) The flow-matching based speech generation framework including a text encoder, duration adaptation module, and flow-prediction network. 
 }
 \label{fig:whole}
\end{figure*}
\section{Related Works}
In this section, we briefly introduce related studies to set the stage for our research and highlight the novelty of our contributions. We first discuss the hierarchical structure of speech emotions, which motivates the use of multi-level representations to better capture emotional nuances. We review related studies on prosody modeling and emotion intensity control in TTS, both of which aim to synthesize speech that varies naturally in pitch, rhythm, and emotional strength. Finally, we highlight how diffusion models have recently been integrated into TTS, inspiring our use in this research.

%followed by an overview of related topics in TTS, including prosody modeling, emotion intensity control, and TTS with diffusion models.

\subsection{Speech Emotion and Its Hierarchical Structure} \label{sec:speech_emotion}
%Speech emotions exhibit a hierarchical nature, and we have summarized their characteristics at three different levels: utterance, word and phoneme.

%At the utterance level, emotions are typically reflected through global prosodic patterns such as pitch features (including contour, range, mean, and intonation), tempo, and rhythm~\cite{hy-utt}.
%At the word level, lexical information contributes to the emotional tone of a word \cite{hy-word1}, where the emphasis on words with emotional valence can amplify emotional intensity \cite{hy-word2}. A study \cite{hy-word3} also shows that listeners are more likely to rely on the prosodic cues when there is a conflict between perceived lexical and prosodic emotions. Therefore, it is necessary to assign unexpected emotions to words for comprehensive emotion control.
%At the phoneme level, emotion is expressed through the prosodic features of individual speech sounds like pitch, energy, and duration~\cite{hieprosody}. Various studies demonstrated the existence of emotion at the phoneme level~\cite{hy-ph1,hy-ph2,hy-ph3}.
Speech attributes exhibit a hierarchical nature~\cite{rlaccent,hieprosody,hy-utt,hy-word1}, with emotional expressions characterized at the utterance, word, and phoneme levels.
%Speech emotions manifest in a hierarchical nature, characterized by distinct features at the utterance, word, and phoneme levels. 
At the utterance level, emotions are typically conveyed through global prosodic patterns such as pitch features (including contour, range, mean, and intonation), tempo, and rhythm~\cite{hy-utt}. At the word level, lexical information contributes to the emotional tone of a word~\cite{hy-word1,rlsentiment}, where emphasis on words with emotional valence can amplify emotional intensity \cite{hy-word2}. A study \cite{hy-word3} shows that listeners are more likely to rely on prosodic cues when there is a conflict between perceived lexical and prosodic emotions, highlighting the need to assign unexpected emotions to words for comprehensive emotion control. At the phoneme level, emotion is expressed through the prosodic features of individual speech sounds, such as pitch, energy, and duration~\cite{hieprosody}. Various studies have demonstrated the existence of emotion at the phoneme level~\cite{hy-ph1,hy-ph2,hy-ph3}. This hierarchical emotion structure underscores the importance of considering multi-level emotional information for effective emotion modeling and control.

%Since speech emotion possesses distinct features at the utterance, word, and phoneme levels, this hierarchy underscores the necessity of considering multi-level emotion information for effective emotion modeling and control.

\subsection{Prosody Modeling in TTS}

%The burgeoning interest in expressive Text-to-Speech (TTS) systems focuses on enhancing speech synthesis through the integration of prosodic features. 
Research in expressive Text-to-Speech (TTS) synthesis has demonstrated the effectiveness of multi-level prosody modeling in enhancing speech expressiveness. Some studies utilize a reference encoder, such as Global Style Tokens (GST) \cite{gst} and Emotion-enhanced GST \cite{emogst} to improve emotion-related prosody modeling. These frameworks have been further advanced through semi-supervised training to link GST tokens with emotion labels~\cite{wu2019endtoend}, alongside other methods for improving expressive TTS quality~\cite{he2022improve, 9413398, yoon2022language}. Particularly, He et al.\cite{he2022improve} addressed the limitation of emotional dataset scarcity through semi-supervised training of graph neural networks. %In \cite{9413398} advanced speaker disentanglement by implementing an emotion information bottleneck.
Several studies adopt multi-level prosodic features to enhance the performance of TTS. For example, phone-level content-style disentanglement~\cite{Tan_2021} and multi-resolution VAEs~\cite{hono2021hierarchical} have been successfully applied to generate multi-scale style embeddings, which are then integrated into TTS models based on VAE~\cite{li2021multiscale}, Vision Transformers (VITs) \cite{kim2021conditional,zhao2023article}, or Diffusion with GANs~\cite{goodfellow2014generative} to accelerate the decoding process, FastDiff~\cite{huang-etal-2023-fastdiff}, which employs a noise scheduling algorithm~\cite{lam2022bddm} to reduce steps, and MatchaTTS~\cite{mehta2024matchatts} which utilizes optimal-transport conditional flow matching (OT-CFM)~\cite{lipman2023flow} to enhance synthesis speed, aim to improve efficiency. 
In our experiments, we adopt MatchaTTS as the backbone model due to its lightweight framework design and state-of-the-art performance among neural TTS systems.

\section{Methodology}
%\textcolor{red}{[Hi Sho, you need to understand the difference between main section and experimental setup. Only in experimental setup, we give all training details, model implementation details and dataset division. Main section should be focused on your contributions and the novel design, and explain the reason why you did that. The main section is not a ``technical report". If you take a look back at your ICASSP paper, I am sure you will understand.]}

%\textcolor{red}{Add here one paragraph to briefly talk about the overall work flow}

Our proposed framework comprises a text encoder, a speaker encoder, a hierarchical emotion distribution (ED) extraction module, and a flow-prediction network, as illustrated in Fig.\ref{fig:whole}~(b,c). 
%Our proposed system comprises four main steps. 
%Initially, we train speech emotion classification models using the waveform data (Fig.\ref{fig:training}). Subsequently, we utilize these trained emotion classifiers to extract the hierarchical emotion distribution (ED) (Fig.\ref{fig:emotion_intensity}). Following this, we train emotion text-to-speech (TTS) models, guiding the emotional output based on the hierarchical ED (Fig.~\ref{fig:whole}). During inference, we allow for manual adjustments to the emotion intensities across three different levels.%\subsection{Problem Formulation}
Given an audio input and its transcript, the hierarchical emotion distribution (ED) extraction module produces a hierarchical ED, which can be modified according to the user's intentions during inference. 
%we aim to develop a model that can modulate the emotional intensity of speech segments. To achieve this, we design a hierarchical Emotion Distribution (ED) extractor that produces a hierarchical ED, which can be adjusted according to the user's intentions. 
The speaker encoder, built using Resemblyzer\footnote{Resemblyzer: \url{https://github.com/resemble-ai/Resemblyzer}}~\cite{resemblyzer}, extracts speaker embeddings. The text encoder generates linguistic embeddings from the input text. Together with the speaker and hierarchical ED embeddings, these embeddings serve as conditions for the flow-prediction network, which iteratively predicts the mel-spectrogram using a diffusion approach. 
%During training, the TTS framework in learning a wide range of emotional variations from the reference audio.
%These embeddings condition the speech synthesis process, applicable in both single and mixed-emotion settings across various text-to-speech frameworks. 
%While previous emotional TTS frameworks rely on utterance-level emotions without detailing intensity variations, our approach 
During TTS training, the hierarchical ED serves as `soft labels' to the flow-prediction model, guiding the model to learn fine-grained emotional information in a quantifiable and hierarchical manner.
%While typical labeling in speech databases focuses on utterance-level emotions without detailing intensity variations, our approach automatically generates multi-level and quantitative intensity labels, serving as `soft labels' for TTS models, thereby eliminating the need for manual emotional intensity labeling.

\begin{figure*}[t]
  \centering
  \centerline{\includegraphics[width=17.6cm]{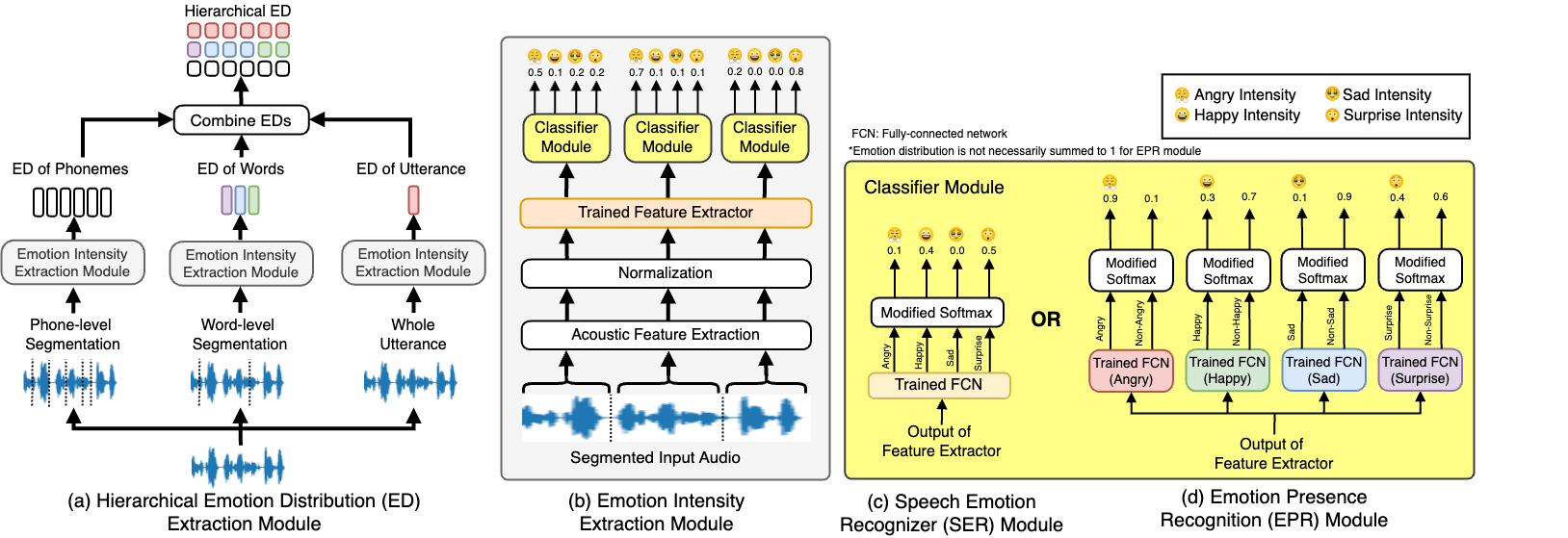}}
  %\vspace{-0.3cm}
 \caption{
 Diagrams of (a) Hierarchical ED Extraction Module; (b) Emotion Intensity Extraction Module; (c) SER classifier module; (d) EPR classifier module. The hierarchical ED extraction module consists of three identical emotion intensity extraction modules, which are applied to the phoneme, word, and utterance levels, respectively.
}
 \label{fig:emotion_intensity}
\end{figure*}

\begin{figure}[t]
  \centering
  \centerline{\includegraphics[width=8.5cm]{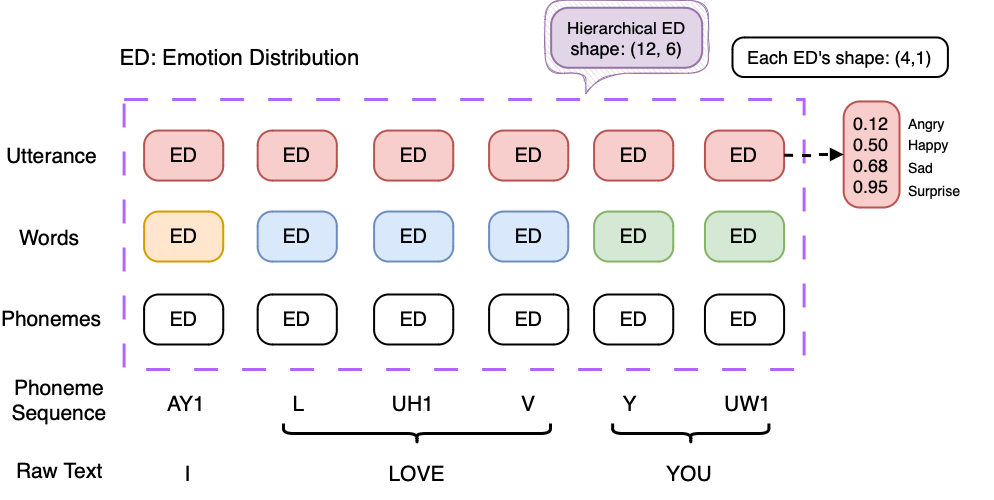}}
  %\vspace{-0.3cm}
 \caption{
 An example of hierarchical emotion distribution (ED), composed of EDs at levels of utterance, words, and phonemes.
}
\vspace{-0.5cm}
 \label{fig:hed}
\end{figure}

\subsection{Hierarchical Emotion Distribution (ED) Extraction}
%\textcolor{blue}{Sho changed the content a bit to adapt to the name change in the figure~\ref{fig:emotion_intensity}}.

%\textcolor{red}{Hi Sho, you need to specify 1) Hierarchical Emotion Distribution Extraction and 2) Emotion Intensity Extraction. I find these two terms are confusing.  }
%\textcolor{red}{Hi Sho, in Fig.2, I can't find "emotion intensity extraction" that described in Fig4. Fig 2 is the overall diagram which should include all essential components.}

The hierarchical ED extraction module comprises speech segmentation and three identical emotion intensity extraction modules for each segment, as illustrated in Fig.\ref{fig:emotion_intensity}~(a). 
Given input audio and transcription, the hierarchical ED extraction module yields the emotion distribution for any segment. The extraction module is applied at the phoneme, word, and utterance levels, with the resulting features concatenated to create the hierarchical ED (Fig.\ref{fig:hed}).
%\textcolor{green}{By applying the emotion intensity extractor to three segments, we are able to extract the hierarchical ED.}
%The diagram demonstrates the emotion intensity extraction in input audio with three words. 
As shown in Fig.\ref{fig:emotion_intensity}(b), the emotion intensity extraction module comprises an acoustic feature extraction module, a normalization layer, a trained feature extractor, and a classifier module.
%\textcolor{red}{I am not clear about "emotion classification". Which part in Fig 4 corresponds to it?}
%We delve into each module as follows. 
The acoustic features are derived from the raw input audio using the acoustic feature extraction modules such as OpenSMILE~\cite{opensmile} or self-supervised learning (SSL) encoders (like WavLM~\cite{WavLM} and Hubert~\cite{Hubert}). 
%\textcolor{red}{Can you merge Sec.III-E here? You cannot let readers to refer to other sections. Every section itself should be clear enough.} 
Acoustic features are standardized across audio samples in the training dataset. The feature extractor comprises two fully connected layers separated by a ReLU activation layer. After feature extraction, the classifier module determines the classification probability, which we regard as emotion intensity.

We develop two types of classifier modules: Speech Emotion Recognizer (SER) and Emotion Presence Recognizer (EPR). The SER, a single fully-connected network (FCN), predicts emotion and has an output size of four, corresponding to the four emotions. In contrast, each FCN in the EPR predicts whether the input corresponds to a specific emotion. For instance, the red FCN in Fig.\ref{fig:emotion_intensity}(d) identifies whether the input is Angry or Non-Angry. We adjusted the softmax function's temperature to prevent excessive confidence in emotion prediction, which skews probability values close to 0.0 or 1.0. We employ a modified Softmax function, denoted as $s(z_i)=\frac{\alpha^{z_i}}{\sum^K_{j=1}\alpha^{z_j}}$, where $\alpha$ is the constant value to control entropy in the softmax distribution, $i, j\in \{0,1,...,K-1\}$ are the indices of the output layer's nodes, and $K$ is the number of classes (emotions). 
Varying $\alpha$ from 1.1 to 3.0 in 0.1 increments, we calculated the KL divergence between the uniform distribution and the emotion intensity distribution of training samples. We then selected the $\alpha$ with the lowest KL divergence score to smooth the distribution of emotion intensities. For each audio sample, we apply emotion intensity extraction at utterance, word, and phoneme levels. 
\textcolor{black}{
This $\alpha$ range meaningfully modifies the softmax function (since $\alpha=1.0$ yields no change) and mitigates the overconfidence inherent in the original exponential function (with base $e\approx2.718$). 
}
The combination of these intensities is shown in Fig.\ref{fig:hed}. The utterance-level intensity is replicated to phoneme length, and the word-level one is adjusted to match the corresponding phonemes.

\begin{figure}[t]
  \centering
  \centerline{\includegraphics[width=8.2cm]{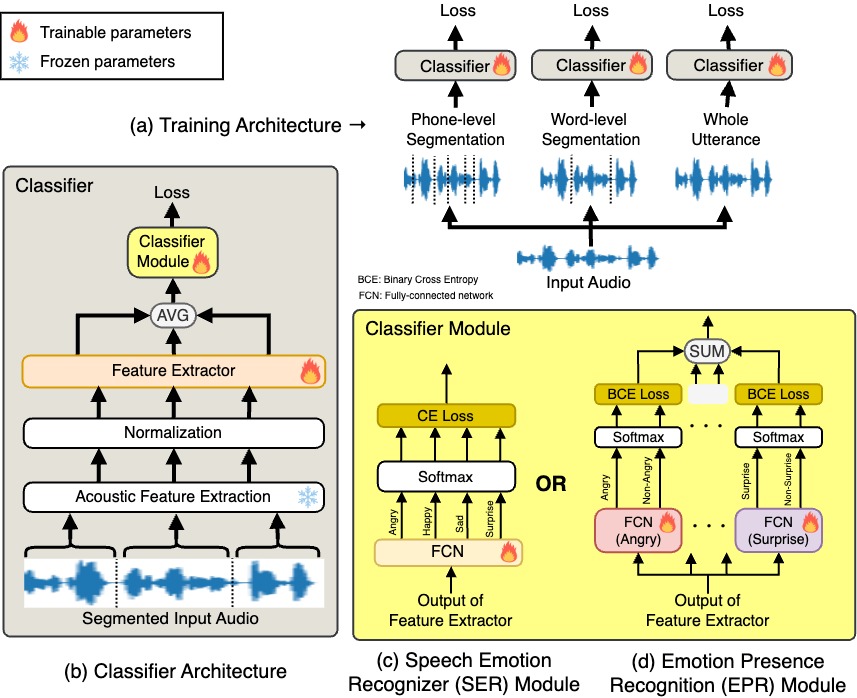}}
 \caption{
 Training diagrams emotion intensity extractors (a) Whole diagram; (b) Classification for each segment; (c) SER classifier module; (d) EPR classifier module.
}
 \label{fig:training}
\end{figure}

\subsection{Training Procedure}

Fig.\ref{fig:training} presents the training diagram for the emotion intensity model. As illustrated in Fig.\ref{fig:training}(a), we segment each audio sample into utterance, word, and phoneme segments, with a single model classifying all segments. During training, as depicted in Fig.~\ref{fig:training}(b), we freeze the parameters of the acoustic feature extraction module and train only the feature extractor and classifier modules.
The model averages the feature extractor's outputs and classifies them using either the SER (Fig.~\ref{fig:training}(c)) or EPR (Fig.~\ref{fig:training}(d)) modules. 
We then extracted the hierarchical emotion distribution by segmenting and applying our trained emotion intensity extractor.

In TTS training, reference audio and its transcription serve as inputs. From the audio, we extract hierarchical ED and speaker embedding, which are then concatenated with linguistic embedding. This embedding is expanded based on phoneme-level duration to compute the predicted average mel-spectrogram, $\mu$. This $\mu$ conditions the diffusion decoding process in a flow-prediction network~\cite{mehta2024matchatts}, taking inputs from flow-matching steps $t \in [0,1]$ and datum $x_t$; where $x_0 \sim N(0, I)$ and $x_1$ represents the target mel-spectrogram (Fig.\ref{fig:whole}(c)).
  
\subsection{Hierarchical Emotion Control}

During inference, our framework supports hierarchical, user-driven control over emotion rendering in any given utterance (``emotion transfer"), as illustrated in Fig.~\ref{fig:whole}(a), one can increase the intensity of a ``Sad" emotion for a single phoneme, thereby achieving fine-grained emotion control. 

Given any unseen $\{$audio, text$\}$ inputs, we begin by extracting a hierarchical emotion distribution (ED) from the input audio. Building on the hierarchical emotion modeling principles, our framework allows users to adjust the ED at three segmental levels---phoneme, word, and sentence---to fine-tune emotional rendering. These user modifications to the ED are then combined with linguistic embeddings from the text encoder, guiding the generative process within our flow-matching based framework. Conditioned on the average mel-spectrogram $\mu$, the flow-prediction network iteratively refines a noise-initialized sample into a mel-spectrogram that reflects the user’s chosen emotional adjustments. In the experimental sections, we will demonstrate how these manual, hierarchical manipulations translate into practical, quantifiable emotional changes in the synthesized speech, showcasing the flexibility and effectiveness of our approach.

%Users can adjust emotion distributions at three levels to modify rendered emotions. For example, Fig.~\ref{fig:whole}(a) shows an increase in the ``Sad'' intensity of the first phoneme. The modified hierarchical ED is then concatenated with the linguistic embedding from the text encoder. The model conditions the diffusion process on the average mel-spectrogram $\mu$. The flow-prediction network iteratively predicts the derivative $\frac{d}{dt}\phi_t(x)$ of the flow $\phi_t$, where $t$ indicates the flow-matching step. Starting from $x_0 \sim N(0, I)$, it generates the emotion-modified mel-spectrogram $x_1$.

\section{Experiment Setup}

We conducted all the experiments with the Emotional Speech Dataset (ESD)~\cite{zhou2021seen, esd}, which consists of over 29 hours of Chinese and English emotional speech categorized into five emotions: Neutral, Angry, Happy, Sad, and Surprise. We exclusively use the English subset from ESD, all sampled at 16 kHz. For each speaker and emotion, we used 300, 20, and 30 speech samples for the training, validation, and test datasets, respectively, following the division used in the ESD dataset. 
%\textcolor{red}{I don't really understand the diffusion generation pipeline, can you introduce more details here?}\textcolor{blue}{wait a moment, let me put the detail in the overall figure}
We sampled random noise $x_0$ from the Gaussian distribution several times to generate speech samples under the same condition.
We generated five samples per test case from 200 samples in the ESD dataset for experiments on speech intelligibility and expressiveness, resulting in 1000 evaluation samples. For other experiments, each of the 50 samples was generated twice, totaling 100 samples.

\subsection{Model Architecture}

We adopted MatchaTTS~\cite{mehta2024matchatts} as our backbone TTS framework, which consists of a text encoder, a duration predictor, and a decoder. The text encoder, built on a transformer network~\cite{transformer}, converts phoneme sequences into linguistic embeddings. The duration predictor, comprising two convolutional layers followed by a projection layer, estimates the required frame count for each phoneme. The decoder is a U-Net architecture with 1D convolutional residual blocks that perform both downsampling and upsampling, with a Transformer block following each residual block. The TTS system is trained using optimal-transport conditional flow matching (OT-CFM)~\cite{lipman2023flow}. We also made two key modifications to the original MatchaTTS: (1) phoneme alignment was derived from Montreal Forced Alignment (MFA) \cite{kim2020glowtts} to reduce computational demand; and (2) the decoder's parameter set was increased from 16M to 160M to accommodate emotion conditioning, multiple speakers, and an expanded output of mel-spectrogram dimension from 80 to 100. 
%Furthermore, we enhanced conditioning capabilities using classifier-free guidance (CFG), where we utilize the duration information from the conditional prediction for the unconditional prediction. The guidance scale is defined by the product of the constant $\lambda$ and the gradient ratio of unconditional to conditional score functions~\cite{kim2022guidedtts}.

%\textcolor{red}{What is the input to the framework? You need to describe the input process: extract 100-d mel with xxx FFT size ....} 

To enable multi-speaker scenarios, we integrated speaker embeddings into the output of the linguistic encoder, using Resemblyzer~\cite{wan2020generalized} for speaker encoding. 
To enhance speaker disentanglement during training, we extracted speaker embeddings from different audio samples corresponding to the same speaker as in the reference audio.
%\textcolor{red}{What does "varied embeddings from the same speaker" mean?} 
A fully connected layer was then used to process these speaker embeddings along with hierarchical emotion distributions, ensuring alignment with the text encoder's output size. For vocoding, we employed Vocos~\cite{siuzdak2023vocos}, which synthesizes audio signals from mel-spectrograms. 
The Mel-spectrogram is configured with 100 dimensions, an FFT size of 1024, and a hop length of 256. The audio samples were downsampled to 16kHz.
%, matching the ESD dataset's sampling rate~\cite{esd} for vocoder training. \textcolor{red}{I don't understand here. Where are those audio samples from? I think you mentioned that "we conduct all experiments with ESD", why do they need to be downsampled?}

\subsection{Training Configuration}

We first trained classifiers for the hierarchical Emotion Distribution (ED) extraction module and then used the extracted hierarchical ED to train the emotion Text-to-Speech (TTS) system.  
For the emotion classifier, we employed the Adam optimizer~\cite{adam} with an initial learning rate of 0.001, set the batch size to 16, and utilized a scheduler that reduces the learning rate by 0.8 every five steps. The loss function used was cross-entropy, with loss weights adjusted based on the segment count.
%\textcolor{red}{Did you first train classifier then train the TTS? You need to make it clear.} 

To remove the speaker and the gender information from the emotion intensity, we integrated an adversarial classifier with a Gradient Reversal Layer (GRL)~\cite{ganin2015unsupervised,gaogrlemo}. 
%This classifier uses the averaged outputs from the feature extractor to predict the speaker or gender. We set the GRL's scaling parameter to 0.5. 
{
The classifier receives the averaged outputs from the shared feature extractor to predict speaker or gender labels. By reversing its gradients (scaled by 0.5), the extractor is trained to suppress speaker- and gender-related cues, promoting disentangled representations. This enhances speaker invariance in our multi-level emotion intensity framework, allowing both the Emotion Presence Recognizer (EPR) and Speech Emotion Recognizer (SER) to better focus on emotional content.
}
We selected the model that achieved the highest validation accuracy in emotion classification and approached random classification accuracy (e.g., 0.2 for a five-label classification) in speaker/gender prediction. To ensure stability in GRL training, we fixed the feature extractor and trained only the classifier for 100 epochs, followed by an evaluation of the validation datasets.

\subsection{Model Comparison}
%\textcolor{red}{I think you can move section 3.D to here}
%\textcolor{red}{Here you should briefly describe the two baseline models}

%\subsection{Comparison with different acoustic representations}\label{sec:acoustic_features}
%\textcolor{red}{I think subsection D should be moved to experiment}
%In this paper, we employ various acoustic features ranging from simple ones like OpenSMILE~\cite{opensmile} to more sophisticated self-supervised learning (SSL) features, such as Hubert~\cite{hsu2021hubert}, WavLM~\cite{Chen_2022}, and emotion2vec~\cite{ma2023emotion2vec}. OpenSMILE offers a comprehensive set of predefined features specifically designed to capture essential prosodic and spectral characteristics for speech and emotion analysis. Among these, we utilized eGeMAPSv02\footnote{OpenSMILE: \url{https://audeering.github.io/opensmile-python/}}, which consists of an 88-dimensional audio set. This set served as a baseline model~\cite{ShoICASSP} in this paper.

We provide a thorough comparison between various baseline models and the proposed models under different configurations. We incorporated MsEmoTTS~\cite{msemotts} (``Baseline MsEmoTTS") and SVM-based Hierarchical Emotion Distribution~\cite{ShoArxiv,ShoICASSP} (``SVM-based HED") into MatchaTTS~\cite{mehta2024matchatts} as baseline models. 
Note that both baseline models utilize OpenSMILE to extract 88-dimensional acoustic features and relative attributes~\cite{parikh2011relative} for quantifying emotion intensity. ``Baseline MsEmoTTS" uses a global emotion label and phoneme-level intensity as conditioning for the backbone TTS framework, whereas ``SVM-based HED" employs a hierarchical emotion distribution across phone, word, and utterance levels. 

As for our proposed models, we first compare two classifier modules settings: Speech Emotion Recognizer (SER) (``Proposed w/ SER") and Emotion Presence Recognizer (EPR) (``Proposed w/ EPR"). In terms of acoustic features, the proposed framework combined OpenSMILE and WavLM to enhance both global and local representations. Specifically, WavLM was employed to extract utterance-level emotion distribution, and OpenSMILE was used for words and phonemes.

\section{Results}
\begin{table*}[t]
\caption{
Objective evaluation result for speech intelligibility, expressiveness, and speaker similarity: MCD, Pitch, and Energy denote mel cepstral, pitch, and energy distortions, respectively. The green and the red colors indicate the best and the worst values among models, respectively.
}
\label{table:objective_result}
\begin{center}
\begin{tabular}{lcccccc}
\toprule
 & \multicolumn{1}{c}{WER ($\downarrow$)} & \multicolumn{3}{c}{Expressiveness ($\downarrow$)} & \multicolumn{2}{c}{SECS ($\uparrow$)} \\
\cmidrule(lr){2-2}\cmidrule(lr){3-5}\cmidrule(lr){6-7}
 & Whisper & MCD & Pitch ($10^1$) & Energy ($10^{-2}$) & WavLM & WeSpeaker\\
\midrule
MsEmoTTS (Baseline) \cite{msemotts} &{9.680} &\cellcolor{red!25}6.468{\tiny $\pm$ {0.134} } & \cellcolor{red!25}3.521{\tiny $\pm$ {0.136} } & \cellcolor{red!25}7.290{\tiny $\pm$ {0.628} } & \cellcolor{red!25}0.761 & \cellcolor{red!25}0.221\\
SVM-based HED (Baseline) \cite{ShoICASSP} &{10.21} & 5.314{\tiny $\pm$ {0.126} } & 2.226{\tiny $\pm$ {0.101} } & 7.144{\tiny $\pm$ {0.681} } & 0.865 & 0.488\\
Proposed w/ SER &\cellcolor{red!25}{10.77} & 5.704{\tiny $\pm$ {0.138} } & 2.336{\tiny $\pm$ {0.104} } & 6.119{\tiny $\pm$ {0.562} } & \cellcolor{green!50}0.873 & 0.506\\
Proposed w/ EPR &\cellcolor{green!50}{{8.747}} & \cellcolor{green!50}5.311{\tiny $\pm$ {0.128} } & \cellcolor{green!50}2.175{\tiny $\pm$ {0.106} } & \cellcolor{green!50}5.665{\tiny $\pm$ {0.478} } & 0.871 & \cellcolor{green!50}0.511\\
\bottomrule
\end{tabular}
\end{center}
\end{table*}

\subsection{Objective Evaluation}
We first report the results of objective evaluation to assess the performance in terms of speech expressiveness, emotion controllability, speaker similarity and speech intelligibility. 
\subsubsection{Emotional Similarity}
We assessed emotion similarity by measuring the prosodic differences between the synthesized and the reference audio. We employ three objective metrics: (a) Mel-Cepstral Distortion (MCD)~\cite{mcd,kun-mix} for spectral similarity, (b) Pitch, and (c) Energy Distortion for prosody alignment.
%, and (d) Frame Disturbance (FD)~\cite{fd} for duration deviation. 
%\textcolor{red}{You can't just conclude the experiments with one sentence. Please first introduce the results, analyze the results, and then draw a conclusion.}
The results are summarized in Table~\ref{table:objective_result}. We observe that our proposed framework with emotion presence recognizer (``Proposed w/ EPR") outperformed other models. In the meanwhile, ``Baseline MsEmoTTS" consistently achieves the worst performance among all the frameworks. 
%\textcolor{red}{Do you think we can just delete FD results from the Table? Otherwise you need to explain here why proposed is not good at FD. Reviewer may ask it. and FD is not an very critical metric anyway} 
These results suggest that our proposed framework with emotion presence recognizer (``Proposed w/ EPR") could achieve better performance in terms of the emotional similarity with the reference emotion. 
%These results suggest a consistent trend between prosodic alignment and perceived emotion alignment, with our model performing the best and MsEmoTTS the worst, as shown in the subjective evaluation results.

%\begin{table}[t]
%\caption{Objective Evaluation Result for Speech Expressiveness ($\downarrow$).
%MCD, Pitch, and Energy denote mel cepstral, pitch, and energy distortions, respectively.
%\textcolor{red}{I suggest to remove FD results, and merge Table 1, 3, 4}
%}
%\label{table:mcd}
%\begin{center}
%\scalebox{0.9}{
%\begin{tabular}{lcccc}
%\toprule
% & MCD & Pitch ($10^1$) & Energy ($10^{-2}$) & FD ($10^1$)\\
%\midrule
%MsEmoTTS (Baseline)\cite{msemotts} &\cellcolor{red!25}6.468{\tiny $\pm$ {0.134} } & \cellcolor{red!25}3.521{\tiny $\pm$ {0.136} } & \cellcolor{red!25}7.290{\tiny $\pm$ {0.628} } & \cellcolor{red!25}3.457{\tiny $\pm$ {0.197} }\\
%SVM-based HED (Baseline)\cite{ShoICASSP} &5.314{\tiny $\pm$ {0.126} } & 2.226{\tiny $\pm$ {0.101} } & 7.144{\tiny $\pm$ {0.681} } & \cellcolor{green!50}2.560{\tiny $\pm$ {0.174} }\\
%Proposed w/ SER &5.704{\tiny $\pm$ {0.138} } & 2.336{\tiny $\pm$ {0.104} } & 6.119{\tiny $\pm$ {0.562} } & 2.944{\tiny $\pm$ {0.202} }\\
%Proposed w/ EPR &\cellcolor{green!50}5.311{\tiny $\pm$ {0.128} } & \cellcolor{green!50}2.175{\tiny $\pm$ {0.106} } & \cellcolor{green!50}5.665{\tiny $\pm$ {0.478} } & 2.869{\tiny $\pm$ {0.196} }\\
%\bottomrule
%\end{tabular}
%}
%\end{center}
%\vskip -2.0em
%\end{table}
  
\subsubsection{Emotion Controllability}\label{sec:emotion_controllability}

We evaluated utterance-level emotion controllability with a pre-trained speech emotion recognizer (SER). We trained the SER model~\cite{ser} on the ESD dataset, with data augmentation of  Gaussian noise and time and frequency masking strategy~\cite{ser1,ser2,ser3,ser4,ser5}. 

We used a Speech Emotion Recognition (SER) system to evaluate synthesized audio samples, where we varied the intensity of one target emotion from 0.0 to 1.0 with an increment of 0.2. All other emotions were kept at a constant intensity. To analyze the SER predictions, we calculated three metrics: ``Positive'', ``Negative'', and ``Score''. ``Positive'' represents how well the SER system’s predictions match the intended emotion. It measures the average correlation between the target emotion’s intensity and the SER’s prediction for the same emotion. For example, if we increase the intensity of ``Angry'', the SER should increase its ``Angry'' prediction accordingly. ``Negative'' represents how much the SER system confuses the target emotion with other emotions. We assign zero to any negative correlations to emphasize when the SER makes incorrect predictions. ``Score'' is a final metric we calculate by subtracting the Negative value from the Positive value. It reflects the overall accuracy of the SER in predicting the target emotion while avoiding confusion with other emotions.
%\textcolor{green}{
%We compared our proposed models with the baseline model with the control of utterance-level emotion intensity: ``SVM-based HED''.
%}

As shown in Table~\ref{table:ser_prediction}, the baseline model, ``SVM-based HED'', consistently demonstrates the worst performance across all three metrics. In contrast, the ``Proposed w/ EPR'' model outperforms the others in both the Positive and Negative metrics. This suggests that our proposed approach with EPR provides strong control over emotional expression in synthesized audio, yielding both higher accuracy and fewer errors. Meanwhile, the ``Proposed w/ SER'' model achieves the best performance on the Negative metric, indicating superior handling of misaligned emotions.
%We used a SER to evaluate the emotion classification values (pre-softmax) of synthesized audio samples, varying target emotion intensities from 0.0 to 1.0 in 0.2 increments, while maintaining other emotion intensities unchanged. We computed three metrics from the SER predictions: ``Positive'', ``Negative'', and ``Score''. First, we calculated the correlation between the emotion intensities and the SER outputs. The ``Positive'' metric represents the average correlation values between the same emotion pairs of emotion intensity and the SER prediction (e.g., Angry intensity changes with Angry prediction outputs); the ``Negative'' metric averages misaligned emotion pairs, assigning zero to negative correlations to emphasize mispredictions. The ``Score'' is derived by subtracting ``Negative'' from ``Positive''. According to results in Table~\ref{table:ser_prediction}, our model significantly surpasses the baseline, indicating a high degree of control that aligns with the SER's accurate predictions. Further details on emotion controllability analysis are in Sec.~\ref{sec:analysis_control}.

\begin{table*}[t]
\caption{BWS test results for evaluating emotion controllability at utterance and word levels: The value represents evaluator preferences (\%), with blue and orange indicating the heatmap for audio samples selected as the least and the most expressive, respectively. `Ang', `Hap', `Sad', `Sur', and `Avg' are abbreviations for `Angry', `Happy', `Sad', `Surprise', and `Average', respectively.}
\label{table:main_bws}
\begin{center}
\scalebox{0.95}{
\begin{tabular}{ccc||ccccc|ccccc|ccccc|}
& & & \multicolumn{5}{c|}{SVM-based HED (Baseline)\cite{ShoICASSP}} & \multicolumn{5}{c|}{Proposed w/ SER} & \multicolumn{5}{c|}{Proposed w/ EPR}\\
& & & Ang & Hap & Sad & Sur & Avg & Ang & Hap & Sad & Sur & Avg & Ang & Hap & Sad & Sur & Avg \\
\midrule
\multirow{6}{*}{Utterance-Level} &  & 0.0 & \cellcolor{cyan!48}{47} & \cellcolor{cyan!60}{59} & \cellcolor{cyan!42}{41} & \cellcolor{cyan!90}{88} & \cellcolor{cyan!60}{59} & \cellcolor{cyan!45}{44} & \cellcolor{cyan!39}{38} & \cellcolor{cyan!48}{47} & \cellcolor{cyan!63}{62} & \cellcolor{cyan!49}{48} & \cellcolor{cyan!71}{69} & \cellcolor{cyan!74}{72} & \cellcolor{cyan!63}{62} & \cellcolor{cyan!68}{66} & \cellcolor{cyan!69}{67} \\
 & Least & 0.5 & \cellcolor{cyan!28}{28} & \cellcolor{cyan!12}{12} & \cellcolor{cyan!25}{25} & \cellcolor{cyan!6}{6} & \cellcolor{cyan!18}{18} & \cellcolor{cyan!25}{25} & \cellcolor{cyan!51}{50} & \cellcolor{cyan!35}{34} & \cellcolor{cyan!12}{12} & \cellcolor{cyan!30}{30} & \cellcolor{cyan!19}{19} & \cellcolor{cyan!6}{6} & \cellcolor{cyan!25}{25} & \cellcolor{cyan!19}{19} & \cellcolor{cyan!17}{17} \\
 &  & 1.0 & \cellcolor{cyan!19}{19} & \cellcolor{cyan!22}{22} & \cellcolor{cyan!28}{28} & \cellcolor{cyan!0}{0} & \cellcolor{cyan!17}{17} & \cellcolor{cyan!25}{25} & \cellcolor{cyan!6}{6} & \cellcolor{cyan!12}{12} & \cellcolor{cyan!19}{19} & \cellcolor{cyan!16}{16} & \cellcolor{cyan!6}{6} & \cellcolor{cyan!16}{16} & \cellcolor{cyan!6}{6} & \cellcolor{cyan!9}{9} & \cellcolor{cyan!9}{9} \\
\cmidrule{2-18}
 &  & 0.0 & \cellcolor{orange!18}{16} & \cellcolor{orange!18}{16} & \cellcolor{orange!60}{53} & \cellcolor{orange!3}{3} & \cellcolor{orange!25}{22} & \cellcolor{orange!21}{19} & \cellcolor{orange!13}{12} & \cellcolor{orange!6}{6} & \cellcolor{orange!18}{16} & \cellcolor{orange!14}{13} & \cellcolor{orange!3}{3} & \cellcolor{orange!6}{6} & \cellcolor{orange!6}{6} & \cellcolor{orange!18}{16} & \cellcolor{orange!9}{8} \\
 & Most & 0.5 & \cellcolor{orange!28}{25} & \cellcolor{orange!50}{44} & \cellcolor{orange!28}{25} & \cellcolor{orange!31}{28} & \cellcolor{orange!34}{30} & \cellcolor{orange!21}{19} & \cellcolor{orange!21}{19} & \cellcolor{orange!25}{22} & \cellcolor{orange!25}{22} & \cellcolor{orange!22}{20} & \cellcolor{orange!13}{12} & \cellcolor{orange!38}{34} & \cellcolor{orange!6}{6} & \cellcolor{orange!18}{16} & \cellcolor{orange!19}{17} \\
 &  & 1.0 & \cellcolor{orange!60}{53} & \cellcolor{orange!38}{34} & \cellcolor{orange!18}{16} & \cellcolor{orange!70}{62} & \cellcolor{orange!46}{41} & \cellcolor{orange!63}{56} & \cellcolor{orange!70}{62} & \cellcolor{orange!75}{66} & \cellcolor{orange!63}{56} & \cellcolor{orange!68}{60} & \cellcolor{orange!88}{78} & \cellcolor{orange!60}{53} & \cellcolor{orange!92}{81} & \cellcolor{orange!70}{62} & \cellcolor{orange!78}{69} \\
\midrule
\multirow{6}{*}{Word-Level} &  & 0.0 & \cellcolor{cyan!68}{66} & \cellcolor{cyan!60}{59} & \cellcolor{cyan!74}{72} & \cellcolor{cyan!86}{84} & \cellcolor{cyan!72}{70} & \cellcolor{cyan!93}{91} & \cellcolor{cyan!48}{47} & \cellcolor{cyan!74}{72} & \cellcolor{cyan!100}{97} & \cellcolor{cyan!79}{77} & \cellcolor{cyan!96}{94} & \cellcolor{cyan!63}{62} & \cellcolor{cyan!93}{91} & \cellcolor{cyan!100}{97} & \cellcolor{cyan!88}{86} \\
 & Least & 0.5 & \cellcolor{cyan!28}{28} & \cellcolor{cyan!19}{19} & \cellcolor{cyan!22}{22} & \cellcolor{cyan!6}{6} & \cellcolor{cyan!19}{19} & \cellcolor{cyan!6}{6} & \cellcolor{cyan!25}{25} & \cellcolor{cyan!3}{3} & \cellcolor{cyan!3}{3} & \cellcolor{cyan!9}{9} & \cellcolor{cyan!3}{3} & \cellcolor{cyan!39}{38} & \cellcolor{cyan!9}{9} & \cellcolor{cyan!3}{3} & \cellcolor{cyan!13}{13} \\
 &  & 1.0 & \cellcolor{cyan!6}{6} & \cellcolor{cyan!22}{22} & \cellcolor{cyan!6}{6} & \cellcolor{cyan!9}{9} & \cellcolor{cyan!11}{11} & \cellcolor{cyan!3}{3} & \cellcolor{cyan!28}{28} & \cellcolor{cyan!25}{25} & \cellcolor{cyan!0}{0} & \cellcolor{cyan!14}{14} & \cellcolor{cyan!3}{3} & \cellcolor{cyan!0}{0} & \cellcolor{cyan!0}{0} & \cellcolor{cyan!0}{0} & \cellcolor{cyan!1}{1} \\
\cmidrule{2-18}
 &  & 0.0 & \cellcolor{orange!13}{12} & \cellcolor{orange!21}{19} & \cellcolor{orange!10}{9} & \cellcolor{orange!3}{3} & \cellcolor{orange!12}{11} & \cellcolor{orange!0}{0} & \cellcolor{orange!46}{41} & \cellcolor{orange!13}{12} & \cellcolor{orange!0}{0} & \cellcolor{orange!14}{13} & \cellcolor{orange!0}{0} & \cellcolor{orange!6}{6} & \cellcolor{orange!0}{0} & \cellcolor{orange!0}{0} & \cellcolor{orange!2}{2} \\
 & Most & 0.5 & \cellcolor{orange!13}{12} & \cellcolor{orange!13}{12} & \cellcolor{orange!35}{31} & \cellcolor{orange!28}{25} & \cellcolor{orange!22}{20} & \cellcolor{orange!13}{12} & \cellcolor{orange!13}{12} & \cellcolor{orange!43}{38} & \cellcolor{orange!28}{25} & \cellcolor{orange!25}{22} & \cellcolor{orange!18}{16} & \cellcolor{orange!10}{9} & \cellcolor{orange!21}{19} & \cellcolor{orange!35}{31} & \cellcolor{orange!21}{19} \\
 &  & 1.0 & \cellcolor{orange!85}{75} & \cellcolor{orange!78}{69} & \cellcolor{orange!67}{59} & \cellcolor{orange!81}{72} & \cellcolor{orange!78}{69} & \cellcolor{orange!100}{88} & \cellcolor{orange!53}{47} & \cellcolor{orange!56}{50} & \cellcolor{orange!85}{75} & \cellcolor{orange!73}{65} & \cellcolor{orange!95}{84} & \cellcolor{orange!95}{84} & \cellcolor{orange!92}{81} & \cellcolor{orange!78}{69} & \cellcolor{orange!90}{80} \\
\bottomrule
\end{tabular}
}
\end{center}
\end{table*}

\begin{table}[t]
\caption{Emotion controllability results: The ``Positive'' metric represents the average correlation values between the same emotion pairs of emotion intensity and the SER prediction; The ``Negative'' metric averages misaligned emotion pairs. The ``Score'' is derived by subtracting ``Negative'' from ``Positive''. 
}
%\textcolor{red}{Can you also mention one sentence in the text that why MsEmoTTS is not compared in this experiment?}
\label{table:ser_prediction}
\begin{center}
\scalebox{0.9}{
\begin{tabular}{lccc}
\toprule
 & Positive ($\uparrow$) & Negative ($\downarrow$) & Score ($\uparrow$)\\
\midrule
SVM-based HED (Baseline)\cite{ShoICASSP} &\cellcolor{red!25}0.156 & \cellcolor{red!25}0.036 & \cellcolor{red!25}0.120\\
Proposed w/ SER &0.261 & \cellcolor{green!50}0.007 & 0.254\\
Proposed w/ EPR &\cellcolor{green!50}0.382 & 0.013 & \cellcolor{green!50}0.369\\
\bottomrule
\end{tabular}
}
\end{center}
\end{table}

\subsubsection{Speaker Similarity}
%To condition speaker information, we use speaker embedding to represent the speaker instead of speaker ID with the embedding layer. Therefore, in theory, it can synthesize speech with unmatched speech prompts for the hierarchical ED and speaker. 
%In this experiment, we utilize a different speaker prompt from the prompt for hierarchical ED to assess the speaker information leakage in hierarchical ED extraction methods.
In this experiment, we use pre-trained speaker verification models to assess the speaker similarity between the synthesized and reference speech. We employ two models: WeSpeaker\footnote{WeSpeaker: \url{https://github.com/wenet-e2e/wespeaker}}~\cite{wespeaker} and WavLM Base with X-vector\footnote{\url{https://huggingface.co/microsoft/wavlm-base-plus-sv}}\cite{WavLM}. We then calculate the Speaker Encoding Cosine Similarity (SECS), as reported in Table \ref{table:objective_result}.
%\textcolor{red}{You can't just conclude the experiments with one sentence. Please first introduce the results, analyze the results, and then draw a conclusion.}
We observe that the proposed models outperform the two baseline models in both speaker verification models as shown in Table~\ref{table:objective_result}. The SECS scores for ``Proposed w/ SER" and ``Proposed w/ EPR" show only minor differences, suggesting that the hierarchical emotion distribution (ED) generated by our method contributes to superior speaker disentanglement performance.

%\begin{table}[t]
%\caption{Speaker Encoding Cosine Similarity (SECS)}
%\label{table:secs}
%\begin{center}
%\begin{tabular}{lcccc}
%\toprule
% & \multicolumn{2}{c}{SECS ($\uparrow$)} \\
%\cmidrule(lr){2-3}
% & WavLM & WeSpeaker\\
%\midrule
%MsEmoTTS (Baseline)\cite{msemotts} &\cellcolor{red!25}0.761 & \cellcolor{red!25}0.221\\
%SVM-based HED (Baseline)\cite{ShoICASSP} &0.865 & 0.488\\
%Proposed w/ SER &\cellcolor{green!50}0.873 & 0.506\\
%Proposed w/ EPR &0.871 & \cellcolor{green!50}0.511\\
%\bottomrule
%\end{tabular}
%\end{center}
%%\vskip -2.0em
%\end{table}

\subsubsection{Speech Intelligibility}
We further conduct objective evaluation to assess the intelligibility of synthesized speech. We employ Whisper\footnote{Whisper: \url{https://github.com/openai/whisper}}~\cite{whisper} to generate the transcriptions from the synthesized speech and calculate word error rate (WER) with the
ground-truth transcriptions. For both baseline and proposed models, we derived the emotion representation from the reference audio. We also compare WER using different reference speakers to assess the impact of speaker leakage on speech intelligibility. We present the results in Table~\ref{table:objective_result}. The ``Proposed w/ EPR'' model outperformed other models when using the original speaker. Interestingly, when using different speakers, the WER scores of our model remained comparable to those obtained with the original speaker, unlike the baseline (SVM+HED). This performance underscores the effectiveness of our speaker disentanglement approach in maintaining speech intelligibility across different speakers. 

\begin{table}[t]
\caption{MUSHRA test results, where a higher value represents better performance in naturalness or emotional similarity.
}
\label{table:mushra}
\begin{center}
\scalebox{0.9}{
\begin{tabular}{lcc}
\toprule
 & Naturalness & Emotional Similarity\\
\midrule
Ground Truth &62.23{\tiny $\pm$ {2.710} } & -\\
Vocoder Resynthesis &59.70{\tiny $\pm$ {2.810} } & -\\
\midrule
MsEmoTTS (Baseline) \cite{msemotts} &47.98{\tiny $\pm$ {2.670} } & \cellcolor{red!25}34.26{\tiny $\pm$ {2.760} }\\
SVM-based HED (Baseline) \cite{ShoICASSP} &\cellcolor{red!25}46.92{\tiny $\pm$ {2.320} } & 58.40{\tiny $\pm$ {2.610} }\\
Proposed w/ SER &47.89{\tiny $\pm$ {2.540} } & 55.12{\tiny $\pm$ {2.800} }\\
Proposed w/ EPR &\cellcolor{green!50}48.91{\tiny $\pm$ {2.360} } & \cellcolor{green!50}59.41{\tiny $\pm$ {2.460} }\\
\bottomrule
\end{tabular}
}
\end{center}
\end{table}

%\begin{table}[t]
%\caption{WER Result ($\downarrow$):
%``Diff.~Spk'' represents the scores when we use different speakers from the reference audio. \textcolor{red}{I suggest to remove WER from different speakers. Otherwise it would be difficult to explain why your model works less efficient than the baseline.  }
%%\textcolor{red}{You can't write caption like that. In captions, try to define all the abbreviations, and briefly introduce the metrics. People are supposed to understand the whole table only by reading caption and table without the need to go into the main texts. }
%}
%\label{table:wer}
%\begin{center}
%\begin{tabular}{lcc}
%\toprule
% & WER & WER (Diff. Spk)\\
%\midrule
%Ground Truth &2.667 & -\\
%Vocoder Resynthesis &3.467 & -\\
%\midrule
%MsEmoTTS (Baseline) \cite{msemotts} &9.967 & \cellcolor{green!50}8.254\\
%SVM-based HED (Baseline) \cite{ShoICASSP} &9.933 & \cellcolor{red!25}11.41\\
%Proposed w/ SER &\cellcolor{red!25}10.73 & 11.03\\
%Proposed w/ EPR &\cellcolor{green!50}9.133 & 9.634\\
%\bottomrule
%\end{tabular}
%\end{center}
%\end{table}
\begin{table*}[t]
\caption{
Result of ablation study: Each separated row indicates a different ablation study.
}
\label{table:ablation_objective}
\begin{center}
\scalebox{0.95}{
\begin{tabular}{llccccccc}
\toprule
 &  & \multicolumn{1}{c}{WER} & \multicolumn{3}{c}{Expressiveness} & \multicolumn{2}{c}{SECS} & \multicolumn{1}{c}{SER} \\
\cmidrule(lr){3-3}\cmidrule(lr){4-6}\cmidrule(lr){7-8}\cmidrule(lr){9-9}
 &  & Whisper & MCD & Pitch ($10^1$) & Energy ($10^{-2}$) & WavLM & WeSpeaker & Score\\
\midrule
\multirow{3}{*}{Sec.~\ref{sec:ablation_features}} & Combination &\cellcolor{green!50}8.747 & \cellcolor{green!50}5.311{\tiny $\pm$ {0.128} } & \cellcolor{green!50}2.175{\tiny $\pm$ {0.106} } & \cellcolor{green!50}5.665{\tiny $\pm$ {0.478} } & 0.871 & \cellcolor{green!50}0.511 & \cellcolor{green!50}0.369\\
 & WavLM &\cellcolor{red!25}15.41 & \cellcolor{red!25}5.967{\tiny $\pm$ {0.148} } & \cellcolor{red!25}2.486{\tiny $\pm$ {0.105} } & \cellcolor{red!25}6.911{\tiny $\pm$ {0.617} }  & \cellcolor{red!25}0.867 & 0.508 & 0.314\\
 & OpenSMILE &9.067 & 5.380{\tiny $\pm$ {0.131} } & 2.298{\tiny $\pm$ {0.103} } & 5.877{\tiny $\pm$ {0.499} } & \cellcolor{green!50}0.871 & \cellcolor{red!25}0.506 & \cellcolor{red!25}0.218\\
\midrule
\multirow{3}{*}{Sec.~\ref{sec:ablation_methods}} & EPR &\cellcolor{green!50}9.067 & 5.380{\tiny $\pm$ {0.131} } & 2.298{\tiny $\pm$ {0.103} } & \cellcolor{green!50}5.877{\tiny $\pm$ {0.499} } & \cellcolor{green!50}0.871 & \cellcolor{green!50}0.506 & \cellcolor{green!50}0.218\\
 & SER &9.253 & \cellcolor{red!25}5.725{\tiny $\pm$ {0.149} } & \cellcolor{red!25}2.393{\tiny $\pm$ {0.104} } & 6.135{\tiny $\pm$ {0.551} } & 0.869 & 0.501 & 0.193\\
 & SVM &\cellcolor{red!25}10.21 & \cellcolor{green!50}5.314{\tiny $\pm$ {0.126} } & \cellcolor{green!50}2.226{\tiny $\pm$ {0.101} } & \cellcolor{red!25}7.144{\tiny $\pm$ {0.681} } & \cellcolor{red!25}0.865 & \cellcolor{red!25}0.488 & \cellcolor{red!25}0.120\\
\midrule
\multirow{4}{*}{Sec.~\ref{sec:ablation_GRL}} & EPR w/ GRL &\cellcolor{green!50}8.747 & \cellcolor{green!50}5.311{\tiny $\pm$ {0.128} } & 2.175{\tiny $\pm$ {0.106} } & \cellcolor{green!50}5.665{\tiny $\pm$ {0.478} } & 0.871 & \cellcolor{green!50}0.511 & \cellcolor{green!50}0.369\\
 & EPR w/o GRL&9.787 & 5.318{\tiny $\pm$ {0.135} } & \cellcolor{green!50}2.119{\tiny $\pm$ {0.089} } & 5.804{\tiny $\pm$ {0.505} } & 0.870 & 0.505 & 0.278\\
 & SER w/ GRL &10.77 & \cellcolor{red!25}5.704{\tiny $\pm$ {0.138} } & \cellcolor{red!25}2.336{\tiny $\pm$ {0.104} } & 6.119{\tiny $\pm$ {0.562} }  & \cellcolor{green!50}0.873 & 0.506 & 0.254\\
 & SER w/o GRL&\cellcolor{red!25}11.84 & 5.630{\tiny $\pm$ {0.134} } & 2.325{\tiny $\pm$ {0.104} } & \cellcolor{red!25}6.430{\tiny $\pm$ {0.554} } & \cellcolor{red!25}0.866 & \cellcolor{red!25}0.497 & \cellcolor{red!25}0.207\\
\midrule
\multirow{4}{*}{Sec.~\ref{sec:ablation_SSL}} & C-WavLM &8.747 & 5.311{\tiny $\pm$ {0.128} } & \cellcolor{green!50}2.175{\tiny $\pm$ {0.106} } & \cellcolor{green!50}5.665{\tiny $\pm$ {0.478} } & 0.871 & 0.511 & \cellcolor{green!50}0.369\\
 & C-Hubert &\cellcolor{green!50}8.133 & \cellcolor{green!50}5.295{\tiny $\pm$ {0.127} } & 2.191{\tiny $\pm$ {0.093} } & 5.823{\tiny $\pm$ {0.504} } & \cellcolor{green!50}0.873 & \cellcolor{green!50}0.520 & 0.285\\
 & WavLM &\cellcolor{red!25}15.41 & \cellcolor{red!25}5.967{\tiny $\pm$ {0.148} } & \cellcolor{red!25}2.486{\tiny $\pm$ {0.105} } & \cellcolor{red!25}6.911{\tiny $\pm$ {0.617} }  & \cellcolor{red!25}0.867 & 0.508 & 0.314\\
 & Hubert &11.60 & 5.648{\tiny $\pm$ {0.148} } & 2.322{\tiny $\pm$ {0.110} } & 6.892{\tiny $\pm$ {0.641} } & 0.868 & \cellcolor{red!25}0.507 & \cellcolor{red!25}0.222\\
\bottomrule
\end{tabular}
}
\end{center}
\end{table*}

\begin{table*}[!t]
\caption{BWS test result (ablation study): The value represents evaluator preferences (\%), with blue and orage indicating the heatmap for audio samples selected as the least and the most expressive, respectively. 
 `Ang', `Hap', `Sad', `Sur', `Avg' are abbreviations for `Angry', `Happy', `Sad', `Surprise' and `Average', respectively.}
\label{table:ablation_control}
\begin{center}
\scalebox{0.95}{
\begin{tabular}{ccc||ccccc|ccccc|ccccc|}
& & & \multicolumn{5}{c|}{EPR+OpenSMILE (\ref{sec:ablation_features}\&\ref{sec:ablation_methods})} & \multicolumn{5}{c|}{EPR+WavLM (\ref{sec:ablation_features})} & \multicolumn{5}{c|}{SER+OpenSMILE (\ref{sec:ablation_methods})}\\
& & & Ang & Hap & Sad & Sur & Avg & Ang & Hap & Sad & Sur & Avg & Ang & Hap & Sad & Sur & Avg \\
\midrule
\multirow{6}{*}{Utterance} &  & 0.0 & \cellcolor{cyan!70}{66} & \cellcolor{cyan!82}{78} & \cellcolor{cyan!73}{69} & \cellcolor{cyan!56}{53} & \cellcolor{cyan!70}{66} & \cellcolor{cyan!65}{62} & \cellcolor{cyan!36}{34} & \cellcolor{cyan!86}{81} & \cellcolor{cyan!65}{62} & \cellcolor{cyan!63}{60} & \cellcolor{cyan!56}{53} & \cellcolor{cyan!32}{31} & \cellcolor{cyan!50}{47} & \cellcolor{cyan!93}{88} & \cellcolor{cyan!58}{55} \\
 & Least & 0.5 & \cellcolor{cyan!17}{16} & \cellcolor{cyan!12}{12} & \cellcolor{cyan!20}{19} & \cellcolor{cyan!0}{0} & \cellcolor{cyan!12}{12} & \cellcolor{cyan!23}{22} & \cellcolor{cyan!20}{19} & \cellcolor{cyan!9}{9} & \cellcolor{cyan!20}{19} & \cellcolor{cyan!18}{17} & \cellcolor{cyan!20}{19} & \cellcolor{cyan!17}{16} & \cellcolor{cyan!36}{34} & \cellcolor{cyan!3}{3} & \cellcolor{cyan!19}{18} \\
 &  & 1.0 & \cellcolor{cyan!12}{12} & \cellcolor{cyan!3}{3} & \cellcolor{cyan!6}{6} & \cellcolor{cyan!43}{41} & \cellcolor{cyan!17}{16} & \cellcolor{cyan!9}{9} & \cellcolor{cyan!43}{41} & \cellcolor{cyan!3}{3} & \cellcolor{cyan!12}{12} & \cellcolor{cyan!17}{16} & \cellcolor{cyan!23}{22} & \cellcolor{cyan!50}{47} & \cellcolor{cyan!12}{12} & \cellcolor{cyan!3}{3} & \cellcolor{cyan!22}{21} \\
\cmidrule{2-18}
 &  & 0.0 & \cellcolor{orange!18}{16} & \cellcolor{orange!3}{3} & \cellcolor{orange!13}{12} & \cellcolor{orange!21}{19} & \cellcolor{orange!13}{12} & \cellcolor{orange!13}{12} & \cellcolor{orange!31}{28} & \cellcolor{orange!3}{3} & \cellcolor{orange!18}{16} & \cellcolor{orange!17}{15} & \cellcolor{orange!31}{28} & \cellcolor{orange!50}{44} & \cellcolor{orange!21}{19} & \cellcolor{orange!3}{3} & \cellcolor{orange!26}{23} \\
 & Most & 0.5 & \cellcolor{orange!28}{25} & \cellcolor{orange!31}{28} & \cellcolor{orange!18}{16} & \cellcolor{orange!43}{38} & \cellcolor{orange!30}{27} & \cellcolor{orange!31}{28} & \cellcolor{orange!50}{44} & \cellcolor{orange!13}{12} & \cellcolor{orange!25}{22} & \cellcolor{orange!30}{27} & \cellcolor{orange!35}{31} & \cellcolor{orange!28}{25} & \cellcolor{orange!6}{6} & \cellcolor{orange!53}{47} & \cellcolor{orange!30}{27} \\
 &  & 1.0 & \cellcolor{orange!60}{53} & \cellcolor{orange!70}{62} & \cellcolor{orange!75}{66} & \cellcolor{orange!43}{38} & \cellcolor{orange!62}{55} & \cellcolor{orange!60}{53} & \cellcolor{orange!25}{22} & \cellcolor{orange!88}{78} & \cellcolor{orange!63}{56} & \cellcolor{orange!59}{52} & \cellcolor{orange!38}{34} & \cellcolor{orange!28}{25} & \cellcolor{orange!78}{69} & \cellcolor{orange!50}{44} & \cellcolor{orange!48}{43} \\
\midrule
\multirow{6}{*}{Word} &  & 0.0 & \cellcolor{cyan!100}{94} & \cellcolor{cyan!82}{78} & \cellcolor{cyan!65}{62} & \cellcolor{cyan!93}{88} & \cellcolor{cyan!85}{80} & \cellcolor{cyan!65}{62} & \cellcolor{cyan!79}{75} & \cellcolor{cyan!73}{69} & \cellcolor{cyan!73}{69} & \cellcolor{cyan!73}{69} & \cellcolor{cyan!73}{69} & \cellcolor{cyan!29}{28} & \cellcolor{cyan!89}{84} & \cellcolor{cyan!100}{94} & \cellcolor{cyan!73}{69} \\
 & Least & 0.5 & \cellcolor{cyan!3}{3} & \cellcolor{cyan!20}{19} & \cellcolor{cyan!40}{38} & \cellcolor{cyan!9}{9} & \cellcolor{cyan!18}{17} & \cellcolor{cyan!26}{25} & \cellcolor{cyan!12}{12} & \cellcolor{cyan!17}{16} & \cellcolor{cyan!20}{19} & \cellcolor{cyan!19}{18} & \cellcolor{cyan!23}{22} & \cellcolor{cyan!36}{34} & \cellcolor{cyan!9}{9} & \cellcolor{cyan!6}{6} & \cellcolor{cyan!19}{18} \\
 &  & 1.0 & \cellcolor{cyan!3}{3} & \cellcolor{cyan!3}{3} & \cellcolor{cyan!0}{0} & \cellcolor{cyan!3}{3} & \cellcolor{cyan!2}{2} & \cellcolor{cyan!12}{12} & \cellcolor{cyan!12}{12} & \cellcolor{cyan!17}{16} & \cellcolor{cyan!12}{12} & \cellcolor{cyan!13}{13} & \cellcolor{cyan!9}{9} & \cellcolor{cyan!40}{38} & \cellcolor{cyan!6}{6} & \cellcolor{cyan!0}{0} & \cellcolor{cyan!13}{13} \\
\cmidrule{2-18}
 &  & 0.0 & \cellcolor{orange!0}{0} & \cellcolor{orange!18}{16} & \cellcolor{orange!3}{3} & \cellcolor{orange!0}{0} & \cellcolor{orange!5}{5} & \cellcolor{orange!18}{16} & \cellcolor{orange!6}{6} & \cellcolor{orange!3}{3} & \cellcolor{orange!0}{0} & \cellcolor{orange!6}{6} & \cellcolor{orange!6}{6} & \cellcolor{orange!25}{22} & \cellcolor{orange!10}{9} & \cellcolor{orange!6}{6} & \cellcolor{orange!12}{11} \\
 & Most & 0.5 & \cellcolor{orange!21}{19} & \cellcolor{orange!28}{25} & \cellcolor{orange!10}{9} & \cellcolor{orange!28}{25} & \cellcolor{orange!22}{20} & \cellcolor{orange!21}{19} & \cellcolor{orange!53}{47} & \cellcolor{orange!63}{56} & \cellcolor{orange!25}{22} & \cellcolor{orange!40}{36} & \cellcolor{orange!6}{6} & \cellcolor{orange!43}{38} & \cellcolor{orange!28}{25} & \cellcolor{orange!46}{41} & \cellcolor{orange!30}{27} \\
 &  & 1.0 & \cellcolor{orange!92}{81} & \cellcolor{orange!67}{59} & \cellcolor{orange!100}{88} & \cellcolor{orange!85}{75} & \cellcolor{orange!86}{76} & \cellcolor{orange!75}{66} & \cellcolor{orange!53}{47} & \cellcolor{orange!46}{41} & \cellcolor{orange!88}{78} & \cellcolor{orange!65}{58} & \cellcolor{orange!100}{88} & \cellcolor{orange!46}{41} & \cellcolor{orange!75}{66} & \cellcolor{orange!60}{53} & \cellcolor{orange!70}{62} \\
\bottomrule
\end{tabular}
}
\end{center}
\end{table*}

\subsection{Subjective Evaluation}
We conducted two subjective evaluations: the MUSHRA Test and the Best Worst Scaling (BWS) Test~\cite{bws} with 20 English speakers. Each speaker listens to 386 synthesized samples in total.

\subsubsection{Speech Naturalness and Emotion Similarity}
In our MUSHRA Test, evaluators rated each audio sample from 0 to 100 on speech naturalness and emotion similarity. Participants are asked to focus on the speech naturalness and the emotional style similarity between the reference speech, respectively. 
The Results summarized in Table~\ref{table:mushra} indicate that ``Proposed w/ EPR" outperformed the others in speech naturalness and emotion similarity. 
%This indicates that our novel emotion representation aligns more closely with the reference audio than the baseline representations.

\subsubsection{Emotion Controllability}

We performed the BWS test to assess the controllability of emotion at both word and utterance levels. We randomly selected the audio sample from the test dataset and edited the emotion intensities to three values (0.0, 0.5, 1.0) to synthesize audio samples with low, normal, and high emotion intensities. Then, we instructed human evaluators to identify the least and the most expressive samples among the three. At the word level, we adjusted the intensity for three words with the longest phoneme sequences. We summarized the BWS test results in Table~\ref{table:main_bws}. Our proposed model with EPR produced the most distinguishable outcomes across all four emotions (Angry, Happy, Sad, and Surprise), with listeners consistently identifying the least expressive sample at the lowest intensity and the most expressive at the highest, especially at the utterance level. These findings highlight the effectiveness of our method in controlling emotion intensity across different speech segmental levels.
%These findings underscore the effective alignment of our model's intensity control with perceived human emotions.

\subsection{Ablation Study}

We conducted ablation studies to validate the effectiveness of each component in our proposed system. Table~\ref{table:ablation_objective} displays comprehensive objective evaluation results from these studies, encompassing metrics such as Word Error Rate (WER), speech expressiveness, speaker similarity, and emotion controllability. Additionally, Table~\ref{table:ablation_control} presents results from the Best Worst Scaling (BWS) test, specifically assessing the emotion controllability of models within the ablation study framework. 

\subsubsection{Impact of Acoustic Features}\label{sec:ablation_features}

Acoustic features play a pivotal role in our system, particularly the combined use of OpenSMILE and WavLM in our main experiment, referred to as ``Combination''. We evaluated the effectiveness of ``Combination'' compared to the individual use of ``OpenSMILE'' and ``WavLM''. According to objective evaluations and the BWS results, ``Combination'' outperformed the individual features across most metrics (Tables~\ref{table:ablation_objective}~and~\ref{table:ablation_control}). Interestingly, while ``WavLM'' surpassed ``OpenSMILE'' in utterance-level emotion controllability, ``OpenSMILE" was more effective than ``WavLM'' at capturing word-level perceived emotion intensity. This highlights the synergistic benefit of combining two acoustic features in enhancing the overall system performance.

\subsubsection{Impact of Emotion Intensity Extraction Methods}\label{sec:ablation_methods}

We explored various methods for extracting emotion intensities, including our proposed Emotion Presence Recognition (EPR) model and the Speech Emotion Recognition (SER) model, and compared them against a baseline Support Vector Machine (SVM) model used for hierarchical emotion detection~\cite{ShoArxiv,ShoICASSP}. To ensure a fair comparison, all models utilized the same acoustic feature set, OpenSMILE. Our findings show that ``EPR'' outperforms in metrics such as Word Error Rate (WER), speaker similarity, and utterance-level emotion control. Conversely, ``SVM'' excels in speech expressiveness. In the Best Worst Scaling (BWS) test, both ``EPR'' and ``SER'' significantly outperformed ``SVM'', as documented in Tables~\ref{table:main_bws}~and~\ref{table:ablation_control}.

\subsubsection{Impact of Gradient Reverse Layers}\label{sec:ablation_GRL}

We investigate the use of Gradient Reverse Layer (GRL)~\cite{ganin2015unsupervised}  to enhance speaker disentanglement. We tested 4 different settings in our proposed framework: ``EPR w/o GRL'', ``EPR w/ GRL'', ``SER w/o GRL'', and ``SER w/ GRL''. The results indicated that EPR-based models consistently outperformed SER-based models across all metrics. Notably, the inclusion of GRL generally enhanced the performance of the models, particularly contributing to improvements in the Speaker Emotion Classification Score (SECS). This underscores the effectiveness of GRL in facilitating speaker disentanglement.

\begin{figure*}[t]
\centerline{\includegraphics[width=16cm]{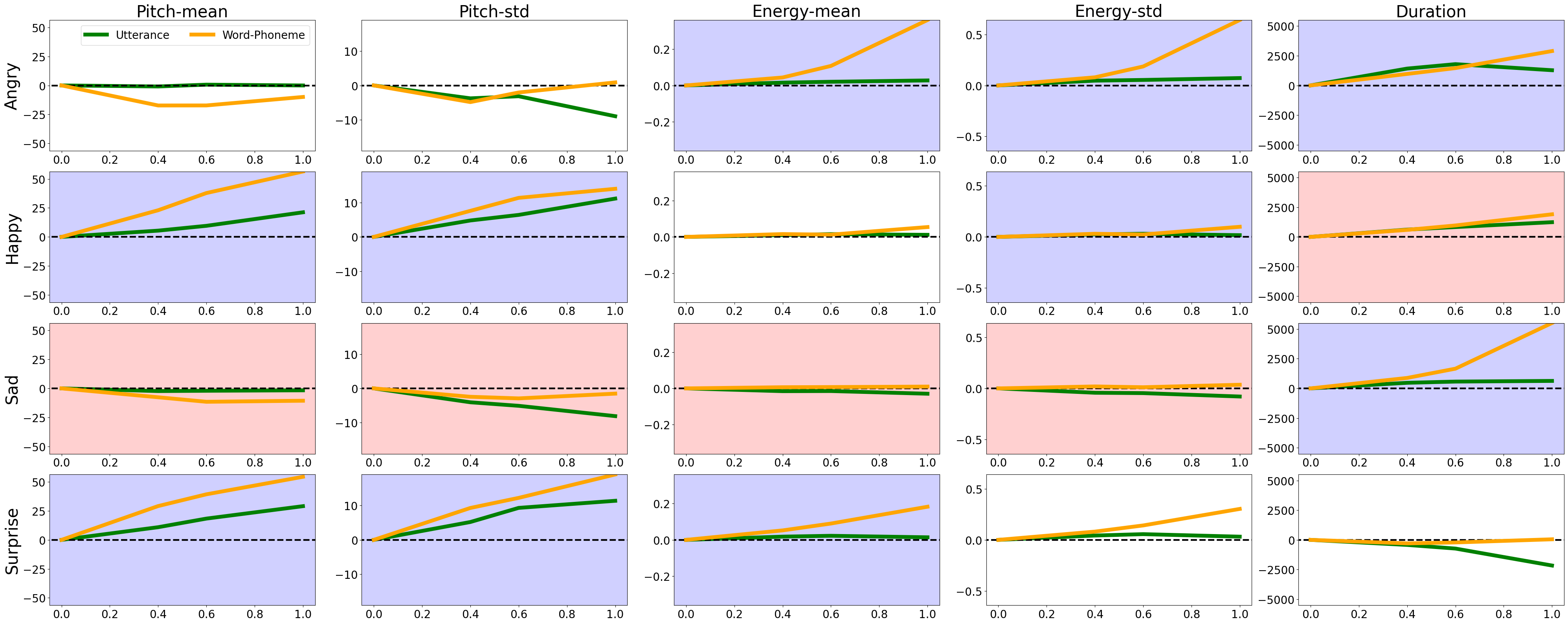}}
\caption{
The illustration of prosodic variants with changes in intensity. A red background indicates the expected negative trend, while blue represents the expected positive trend, as analyzed from the ESD dataset.
}
\label{fig:analysis}
\end{figure*}

\subsubsection{Impact of Different Self-Supervised Learning Features}\label{sec:ablation_SSL}
We further explored the effects of different self-supervised learning (SSL) features including WavLM and Hubert~\cite{Hubert}. Additionally, we evaluated the performance of combining these SSL features with OpenSMILE features, resulting in combinations such as ``C-WavLM" and ``C-HuBERT," similar to the approach used in our proposed methods. In this context, ``C" denotes the integration of emotion distributions derived from each SSL feature at the utterance level, combined with OpenSMILE features applied at the phoneme and word levels.
As shown in Table~\ref{table:ablation_objective}, we observe that SSL features, when combined using our proposed method, consistently outperform those using single features. Specifically, C-HuBERT demonstrates superior performance in WER, SECS, and SER, while C-WavLM excels in capturing emotional expressiveness.
%\textcolor{red}{You need to tell what is the C means, or giving a citation or a definition here} 
Our findings demonstrate that combined acoustic features consistently outperform the individual SSL features. Both WavLM and Hubert showed comparable performance across most metrics. We opted to employ WavLM in our systems primarily due to its superior emotion controllability.

\begin{table*}[!h]
\caption{
\textcolor{black}{
Emotion presence binary classification accuracy across utterance-(U), word-(W), and phoneme-(P) level segments.
}}
\label{table:emotion_presence}
\centering
\scalebox{0.9}{
\begin{tabular}{ccccccccccccccccccc}
\toprule
 \multicolumn{3}{c}{Conditions} & \multicolumn{4}{c}{Angry} & \multicolumn{4}{c}{Happy} & \multicolumn{4}{c}{Sad} & \multicolumn{4}{c}{Surprise}\\
\cmidrule(lr){1-3}\cmidrule(lr){4-7}\cmidrule(lr){8-11}\cmidrule(lr){12-15}\cmidrule(lr){16-19}
 Module & Features & GRL & U & W & P & Avg. & U & W & P & Avg. & U & W & P & Avg. & U & W & P & Avg.\\
\midrule
SVM & OpenSMILE & No & \cellcolor{red!25}{75.2} & \cellcolor{red!25}{67.4} & \cellcolor{red!25}{63.5} & \cellcolor{red!25}{68.7} & \cellcolor{red!25}{71.3} & \cellcolor{red!25}{61.9} & \cellcolor{red!25}{62.2} & \cellcolor{red!25}{65.1} & \cellcolor{red!25}{92.8} & \cellcolor{red!25}{77.6} & \cellcolor{red!25}{69.7} & \cellcolor{red!25}{80.0} & \cellcolor{red!25}{75.4} & \cellcolor{red!25}{67.9} & \cellcolor{red!25}{65.4} & \cellcolor{red!25}{69.6}\\
SVM & WavLM & No & 92.1 & \cellcolor{green!50}{87.4} & \cellcolor{green!50}{82.9} & \cellcolor{green!50}{87.4} & 81.2 & 74.1 & 68.6 & 74.6 & 96.9 & \cellcolor{green!50}{94.4} & \cellcolor{green!50}{90.6} & \cellcolor{green!50}{94.0} & 89.4 & \cellcolor{green!50}{83.9} & \cellcolor{green!50}{79.6} & 84.3\\
\midrule
EPR & OpenSMILE & Yes & 85.9 & 74.7 & 69.6 & 76.7 & 82.1 & 66.4 & 62.3 & 70.3 & 95.1 & 79.7 & 71.4 & 82.1 & 86.9 & 73.1 & 68.9 & 76.3\\
EPR & WavLM & Yes & 95.5 & 82.8 & 76.8 & 85.0 & 91.7 & 75.6 & 71.6 & 79.6 & 97.7 & 88.2 & 82.6 & 89.5 & 93.6 & 79.1 & 75.0 & 82.5\\
EPR & OpenSMILE & No & 89.9 & 76.7 & 71.9 & 79.5 & 87.6 & 71.0 & 67.9 & 75.5 & 94.9 & 81.0 & 72.6 & 82.8 & 89.3 & 74.6 & 70.8 & 78.2\\
EPR & WavLM & No & \cellcolor{green!50}{97.1} & 84.3 & 79.1 & 86.9 & \cellcolor{green!50}{93.7} & \cellcolor{green!50}{77.8} & \cellcolor{green!50}{72.2} & \cellcolor{green!50}{81.2} & \cellcolor{green!50}{98.2} & 88.9 & 84.5 & 90.5 & \cellcolor{green!50}{96.3} & 80.6 & 76.1 & \cellcolor{green!50}{84.3}\\
\bottomrule
\end{tabular}
}
\end{table*}

%\ref{sec:result_emotion_extraction}

\textcolor{black}{
\subsection{Analysis of Hierarchical Emotion Distribution}\label{sec:result_emotion_extraction}
We analyze the hierarchical distribution of emotions by assessing its alignment with ground-truth labels and exploring the disentanglement of speaker characteristics.
}

\textcolor{black}{
\subsubsection{Emotion Classification Accuracy}
We evaluated emotion classifiers through binary classification for emotion presence using SVM-based and EPR-based models. Each classifier determines whether a speech segment conveys a specific emotion. Table~\ref{table:emotion_presence} summarizes the accuracy under different conditions, where ``U'', ``W'', and ``P'' correspond to utterance-, word-, and phoneme-level segments, respectively. Our results demonstrate that acoustic features critically influence performance; WavLM outperforms OpenSMILE. Longer segments yield higher accuracy, although WavLM exhibits a larger drop from utterance- to word- and phoneme-level segments. Additionally, the EPR-based model without GRL performs comparably to the SVM baseline, whereas the EPR with GRL shows reduced accuracy, likely due to adversarial learning of speaker and gender attributes.
}

\textcolor{black}{
Subsequently, we examined the alignment between the hierarchical emotion distribution and the ground-truth labels by computing the ratio of instances in which the emotion with the highest intensity matches the ground-truth label. Table~\ref{table:emotion_accuracy} presents these results, which follow trends similar to those observed in the emotion presence binary classification, with SER values comparable to the EPR models.
}

\begin{table}[!h]
\caption{
\textcolor{black}{Alignment of hierarchical emotion distribution with ground-truth emotion labels across utterance-(U), word-(W), and phoneme-(P) level segment}
}
\label{table:emotion_accuracy}
\centering
\scalebox{0.95}{
\begin{tabular}{ccccccc}
\toprule
 \multicolumn{3}{c}{Conditions} & \multicolumn{4}{c}{Accuracy}\\
\cmidrule(lr){1-3}\cmidrule(lr){4-7}
 Module & Features & GRL & U & W & P & Avg.\\
\midrule
SVM & OpenSMILE & No & \cellcolor{red!25}{68.2} & \cellcolor{red!25}{48.3} & \cellcolor{red!25}{38.0} & \cellcolor{red!25}{51.5}\\
SVM & WavLM & No & 86.8 & \cellcolor{green!50}{77.1} & \cellcolor{green!50}{69.2} & \cellcolor{green!50}{77.7}\\
\midrule
SER & OpenSMILE & Yes & 77.3 & 49.7 & 39.5 & 55.5\\
SER & WavLM & Yes & 89.7 & 62.8 & 52.7 & 68.4\\
SER & OpenSMILE & No & 82.2 & 50.5 & 39.3 & 57.3\\
SER & WavLM & No & 92.3 & 67.1 & 56.6 & 72.0\\
\midrule
EPR & OpenSMILE & Yes & 79.8 & 50.1 & 39.9 & 56.6\\
EPR & WavLM & Yes & 89.9 & 63.6 & 52.9 & 68.8\\
EPR & OpenSMILE & No & 82.3 & 51.4 & 40.6 & 58.1\\
EPR & WavLM & No & \cellcolor{green!50}{92.9} & 66.0 & 55.5 & 71.5\\
\bottomrule
\end{tabular}
}
\end{table}

\textcolor{black}{
\subsubsection{Speaker Disentanglement}
We evaluated speaker disentanglement to quantify speaker leakage in the hierarchical ED, a critical factor for improving emotion representation and controllability. We employed various metrics~\cite{slreview,slmig,sldisentanglement,slexplicitness}.
\\
In the information-based evaluation, we computed the Mutual Information Gap (MIG)~\cite{slmig}. For each factor \(v_i\), we estimated the mutual information with latent code \(z_j\) via
\[
I(v_i, z_j) = \sum_{b_v}\sum_{b_z} P(b_v,b_z) \log\left(\frac{P(b_v,b_z)}{P(b_v)P(b_z)}\right),
\]
where \(b_v\) and \(b_z\) denote bins in the factor and code spaces, respectively. They defined \(I(v_i, z^\star)\) as the highest and \(I(v_i, z^\circ)\) as the second highest mutual information, and computed the MIG as
\[
\text{MIG}(v_i) = \frac{I(v_i, z^\star) - I(v_i, z^\circ)}{H(v_i)},
\]
with \(H(v_i)\) representing the entropy of \(v_i\). The overall MIG score is the average over all factors. In our study, the latent code is derived from a phoneme-level segment of the hierarchical ED with shape \((1,12)\), treating each segment (originally of shape \((\text{\# of phonemes},12)\)) as independent, while the factor corresponds to speaker labels from 10 speakers. We varied the bin count (30, 50, and 100) to ensure reliable estimation.
\\
We computed predictor-based metrics, namely Disentanglement~\cite{sldisentanglement} and Explicitness~\cite{slexplicitness} scores, using Random Forest and Lasso Logistic Regressor classifiers, following the approach in~\cite{sldisentanglement}. We extracted sample-level acoustic features comprising utterance-level emotion intensities and the characteristics of word-level and phoneme-level emotion intensities. For the latter, we computed features including basic statistics (mean, median, standard deviation, maximum, minimum, and interquartile range), linear trend (slope), peak-based metrics (number of peaks and average peak prominence), and first-lag autocorrelation.
\\
Table~\ref{table:speaker_leakage} summarizes the speaker disentanglement results. Lower values indicate superior disentanglement and reduced speaker leakage. The second row in MIG denotes the bin number, while ``RF'' and ``Lasso'' refer to the classification modules. Our proposed models significantly outperform the SVM-based baseline, particularly the EPR-based model. Ablation studies reveal that incorporating a Gradient Reverse Layer (GRL) further enhances speaker disentanglement. 
%These findings suggest that improved speaker disentanglement contributes to enhanced emotion expressiveness and controllability.
}

\textcolor{black}{
We observe a trade-off between emotion controllability and classification accuracy. As shown in Table~\ref{table:emotion_accuracy}, the SVM with WavLM achieves higher accuracy by implicitly leveraging speaker-specific cues. In contrast, our EPR model, equipped with a Gradient Reversal Layer, deliberately suppresses speaker information, as shown in Table~\ref{table:speaker_leakage}. While this leads to a slight drop in classification performance, it significantly reduces speaker leakage and improves control over emotion representations.}

%We clarify that the objectives in Tables~\ref{table:emotion_accuracy} and~\ref{table:speaker_leakage} differ, which explains their divergent results. The SVM with WavLM achieves higher emotion classification accuracy (Table~\ref{table:emotion_accuracy}) by implicitly exploiting speaker cues. In contrast, our EPR model with a Gradient Reversal Layer intentionally suppresses speaker‐specific information, which slightly reduces classification accuracy but substantially lowers speaker leakage and enhances control over emotion representations (Table~\ref{table:speaker_leakage}). This trade‐off is essential for affective speech synthesis, where we must decouple emotion from speaker identity. Thus, the SVM's superior accuracy does not translate into superior emotion modeling for generation, underscoring the importance of disentanglement in ``affective computing''.
%}

\begin{table*}[!h]
\caption{
\textcolor{black}{
Speaker disentanglement evaluation: Mutual information gap (MIG), disentanglement, and explicitness scores for various bin counts and classification modules (``RF'' and ``Lasso''). Lower values denote improved disentanglement and reduced speaker leakage.
}
}
\label{table:speaker_leakage}
\centering
\scalebox{0.95}{
\begin{tabular}{llcccccccccc}
\toprule
 &  & \multicolumn{3}{c}{Condition} & \multicolumn{3}{c}{MIG ($10^{-2}$)~\cite{slmig}} & \multicolumn{2}{c}{Disen. ($10^{-1}$)~\cite{sldisentanglement}} & \multicolumn{2}{c}{Expli. ($10^{-1}$)~\cite{slexplicitness}}\\
\cmidrule(lr){3-5}\cmidrule(lr){6-8}\cmidrule(lr){9-10}\cmidrule(lr){11-12}
  &  & Module & Features & GRL & 30 & 50 & 100 & RF & Lasso & RF & Lasso\\
\midrule
\multirow{3}{*}{Main Models} & Baseline &SVM & OpenSMILE & No & \cellcolor{red!25}{1.843} & \cellcolor{red!25}{1.765} & \cellcolor{red!25}{1.768} & 1.546 & \cellcolor{red!25}{0.785} & \cellcolor{red!25}{7.230} & \cellcolor{red!25}{6.747}\\
 & Proposed w/ SER &SER & C-WavLM & Yes & \cellcolor{green!50}{0.062} & 0.068 & 0.184 & \cellcolor{red!25}{2.530} & \cellcolor{green!50}{0.375} & 5.341 & 5.042\\
 & Proposed w/ EPR &EPR & C-WavLM & Yes & 0.089 & \cellcolor{green!50}{0.03} & \cellcolor{green!50}{0.004} & \cellcolor{green!50}{1.208} & 0.491 & \cellcolor{green!50}{4.784} & \cellcolor{green!50}{4.672}\\
\midrule
\multirow{4}{*}{\shortstack[c]{Sec.~\ref{sec:ablation_GRL}\\(GRL)}} & EPR w/ GRL &EPR & C-WavLM & Yes & 0.089 & \cellcolor{green!50}{0.03} & \cellcolor{green!50}{0.004} & \cellcolor{green!50}{1.208} & 0.491 & \cellcolor{green!50}{4.784} & \cellcolor{green!50}{4.672}\\
 & EPR w/o GRL &EPR & C-WavLM & No & \cellcolor{red!25}{0.530} & \cellcolor{red!25}{0.621} & \cellcolor{red!25}{0.858} & 1.573 & \cellcolor{red!25}{0.803} & \cellcolor{red!25}{5.783} & \cellcolor{red!25}{5.715}\\
 & SER w/ GRL &SER & C-WavLM & Yes & \cellcolor{green!50}{0.062} & 0.068 & 0.184 & \cellcolor{red!25}{2.530} & \cellcolor{green!50}{0.375} & 5.341 & 5.042\\
 & SER w/o GRL &SER & C-WavLM & No & 0.283 & 0.219 & 0.086 & 1.263 & 0.652 & 5.114 & 4.836\\
\bottomrule
\end{tabular}
}
\end{table*}

\section{Discussion}
\subsection{
Analysis of Emotion Controllability
}\label{sec:analysis_control}

We further analyzed our model's controllability over both global and local emotional variations, evaluating it at the utterance and word levels. In line with our previous setup, the emotional intensity of words and phonemes remained consistent. At the word level, we adjusted the emotional intensities for the three words with the longest phonemes, while keeping the utterance intensity constant. We gradually increased the emotion intensity from 0.0 to 1.0, measuring various prosodic features, including duration, pitch (mean and standard deviation), and energy (mean and standard deviation), as illustrated in Fig.\ref{fig:analysis}. These prosodic attributes are strongly correlated with speech emotion, as noted in previous studies \cite{schuller2018speech}. For example, anger often manifests in a slower speaking rate and higher values for energy mean/standard deviation. 

We analyzed the relationship between acoustic features and various emotions, as illustrated in Fig.\ref{fig:analysis} using color to represent positive and negative correlations. A red background indicates an expected negative trend, while a blue background signifies a positive trend with increasing intensity, based on our data analysis from the ESD dataset. For example, anger is typically associated with a slower speaking rate and higher values in both the mean and standard deviation of energy. Our findings show that synthesized emotional speech closely follows the anticipated trends in most prosodic features. Specifically, we observed a positive correlation between happiness intensity and mean pitch, as well as a positive correlation between anger and both the mean and standard deviation of energy. These results demonstrate that our model effectively adjusts acoustic features in response to varying levels of emotional intensity.

\begin{figure}[t]
\centerline{\includegraphics[width=7.5cm]{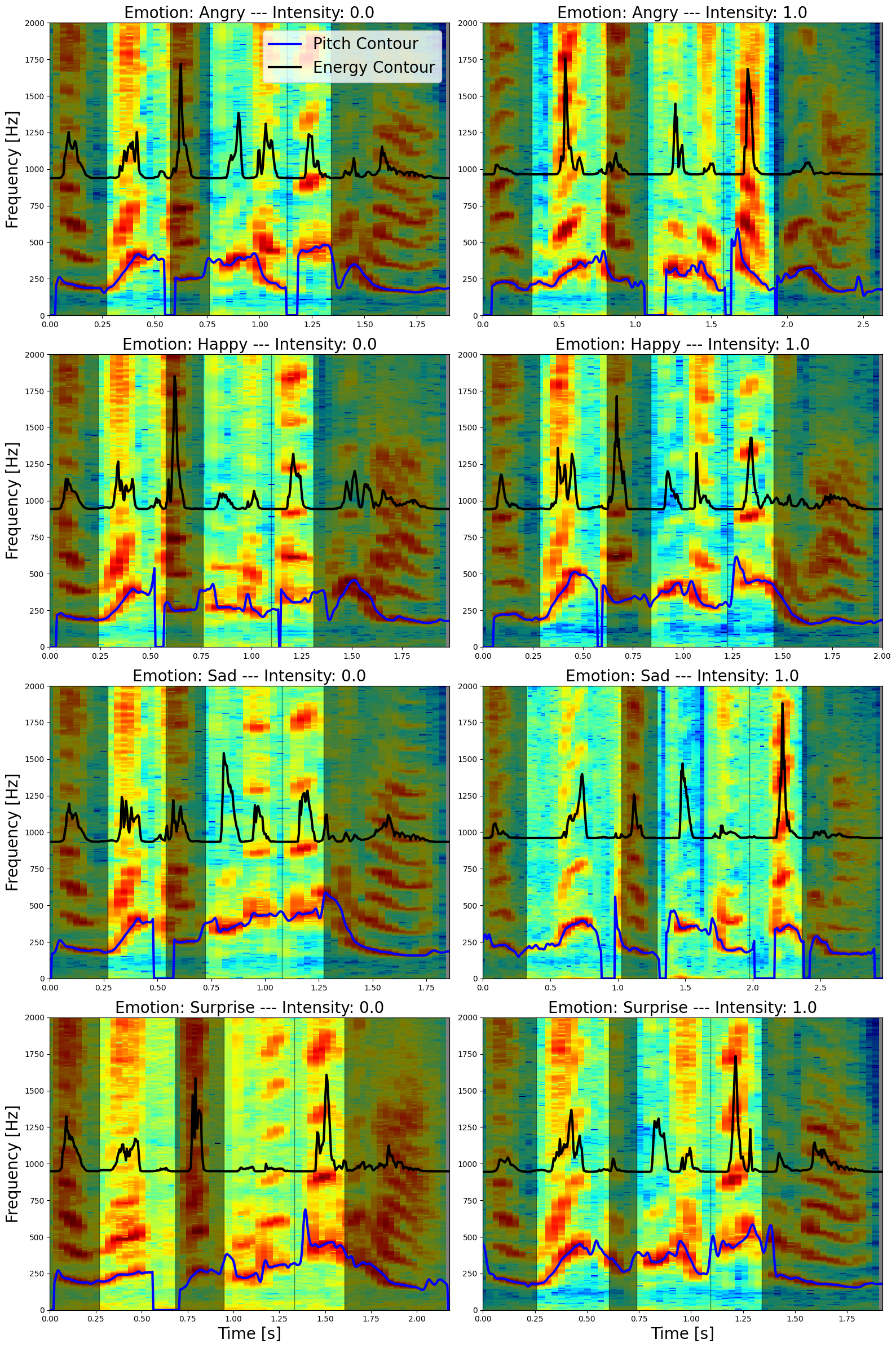}}
\caption{Spectrograms of synthesized audio samples across different emotion intensities with pitch (blue) and energy (black) contours: the y-axis for energy contours are not relevant.}
\label{fig:contours}
\end{figure}

Fig.~\ref{fig:contours} shows the spectrograms of synthesized audio samples with varying emotion intensities. Pitch and energy contours are represented by blue and black lines, respectively. Note that the y-axis for the energy contours is not relevant. Each row corresponds to a different emotion, with the first column depicting acoustic features at an emotion intensity of 0.0, and the second column at 1.0. The three highlighted areas indicate regions where the intensity has been modified. For anger, we observe more pronounced energy spikes, especially in the first and third words, at higher intensities. In happiness, pitch and energy patterns are similar, with higher pitch values at an intensity of 1.0. Sadness is marked by a longer duration and a decline in the pitch contour as intensity increases. For surprise, we note a rise in pitch contours along with a slight energy spike. This demonstrates that our model can effectively manipulate pitch and energy contours in response to changes in emotion intensity.

\subsection{Future Work}

%In future work, we aim to achieve emotion intensity control using naturalistic emotional speech data to better mimic spontaneous emotional expressions. Additionally, we plan to explore state-of-the-art speech emotion recognition and diarization techniques to enhance emotion intensity control. For instance, integrating speech emotion diarization~\cite{EmotionDiarization} could improve frame-level emotion awareness, and extending this research to include multi-modal emotion recognition systems could further advance its applicability~\cite{rlser}.

\textcolor{black}{
Since the dataset used in our study is content-parallel, our hierarchical architecture (at the utterance, word, and phoneme levels) primarily focuses on modeling acoustic dynamics. We acknowledge that lexical content naturally conveys prosodic cues that influence emotional expression, particularly along the valence dimension~\cite{10089511}.
In future work, we plan to further explore the role of linguistic prosody to enhance emotion control. We also aim to extend our model to naturalistic and multi-modal signals (e.g., visual cues~\cite{rlser}), thereby improving its applicability in real-world affective speech synthesis.
}

\section{Conclusion}

In this work, we introduce a flow-matching based emotional text-to-speech framework that offers hierarchical emotion intensity control, enabling precise modulation of emotion rendering at the phoneme, word, and utterance levels. By developing a quantifiable emotion distribution embedding, we provide users with real-time, fine-grained control over emotional rendering during inference. Our investigation of various acoustic features and emotion intensity extractor architectures highlights their significant influence on synthesis quality. Both subjective and objective evaluations demonstrate the effectiveness of our model in achieving high-quality, expressive, and controllable synthesized speech.

%\section*{Acknowledgments}
%This should be a simple paragraph before the References to thank those individuals and institutions who have supported your work on this article.

%\input{sections/appendix.tex}


{\large
\bibliographystyle{IEEEtran}
\bibliography{refs_icassp,refs_spl,refs}
}
%\bibliography{refs}

%\begin{thebibliography}{1}
%
%\bibitem{ref1}
%{\it{Mathematics Into Type}}. American Mathematical Society. [Online]. Available: https://www.ams.org/arc/styleguide/mit-2.pdf
%
%\bibitem{ref2}
%T. W. Chaundy, P. R. Barrett and C. Batey, {\it{The Printing of Mathematics}}. London, U.K., Oxford Univ. Press, 1954.
%
%\bibitem{ref3}
%F. Mittelbach and M. Goossens, {\it{The \LaTeX Companion}}, 2nd ed. Boston, MA, USA: Pearson, 2004.
%
%\bibitem{ref4}
%G. Gr\"atzer, {\it{More Math Into LaTeX}}, New York, NY, USA: Springer, 2007.
%
%\bibitem{ref5}M. Letourneau and J. W. Sharp, {\it{AMS-StyleGuide-online.pdf,}} American Mathematical Society, Providence, RI, USA, [Online]. Available: http://www.ams.org/arc/styleguide/index.html
%
%\bibitem{ref6}
%H. Sira-Ramirez, ``On the sliding mode control of nonlinear systems,'' \textit{Syst. Control Lett.}, vol. 19, pp. 303--312, 1992.
%
%\bibitem{ref7}
%A. Levant, ``Exact differentiation of signals with unbounded higher derivatives,''  in \textit{Proc. 45th IEEE Conf. Decis.
%Control}, San Diego, CA, USA, 2006, pp. 5585--5590. DOI: 10.1109/CDC.2006.377165.
%
%\bibitem{ref8}
%M. Fliess, C. Join, and H. Sira-Ramirez, ``Non-linear estimation is easy,'' \textit{Int. J. Model., Ident. Control}, vol. 4, no. 1, pp. 12--27, 2008.
%
%\bibitem{ref9}
%R. Ortega, A. Astolfi, G. Bastin, and H. Rodriguez, ``Stabilization of food-chain systems using a port-controlled Hamiltonian description,'' in \textit{Proc. Amer. Control Conf.}, Chicago, IL, USA,
%2000, pp. 2245--2249.
%
%\end{thebibliography}

%\newpage

\section{Biography Section}
% your own custom macro containing the $\backslash${\tt{includegraphics}} command to make things
 
\vspace{11pt}

%\bf{If you include a photo:}\vspace{-33pt}
\begin{IEEEbiography}[{\includegraphics[width=1in,height=1.25in,clip,keepaspectratio]{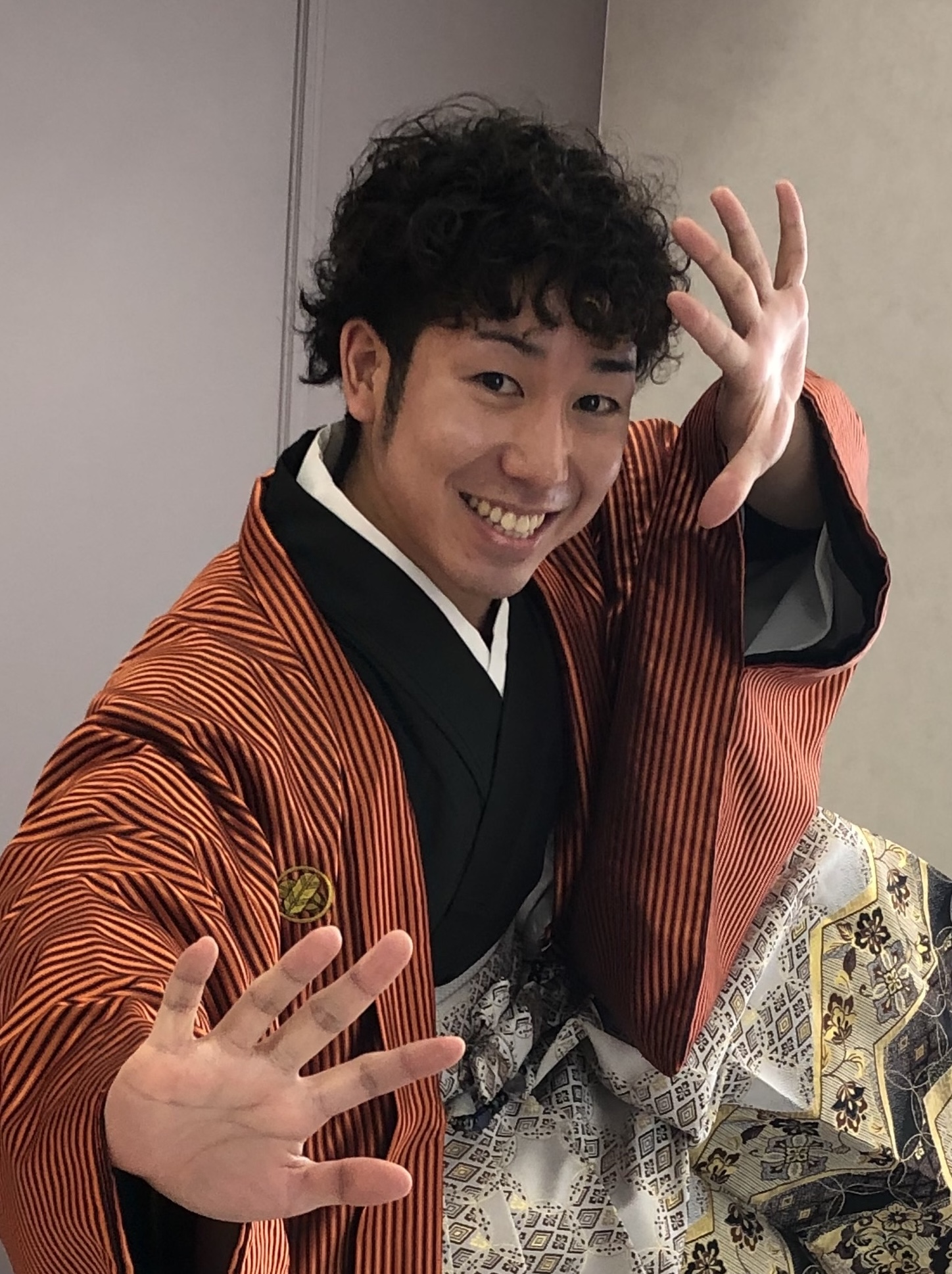}}]{
Sho Inoue % Author name
}
(Student Member, IEEE) received the B. Eng. degree in Electrical Engineering and M.S. in Computer Science at Sophia University, Tokyo, Japan in 2020 and 2022. He is currently a PhD student at the Chinese University of Hong Kong, Shenzhen (CUHK-Shenzhen) in China. His research interests mainly focus on building a speech generation model with free control of paralinguistic features such as emotions, accents, and conversation traits. He has published three conference papers in ICASSP 2024, APSIPA 2024, ICASSP 2025, and Interspeech 2025, with his APSIPA paper being shortlisted for the best paper award. He is also an active reviewer for a leading journal, such as IEEE Transactions on Affective Computing.
\end{IEEEbiography}

\begin{IEEEbiography}[{\includegraphics[width=1in,height=1.25in,clip,keepaspectratio]{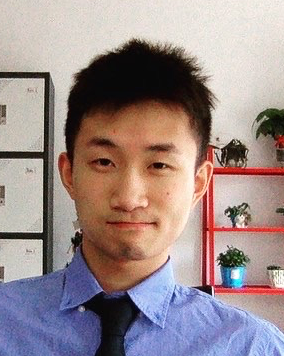}}]{
Kun Zhou % Author name
}(Member, IEEE)
received his B.Eng. degree from the School of Information and Communication Engineering, University of Electronic Science and Technology of China, Chengdu, China, in 2018, and his M.Sc. and Ph.D. degrees in electrical engineering from National University of Singapore, Singapore, in 2019 and 2023, respectively. He is currently a Research Engineer with the Tongyi Speech Team at Alibaba Group. During his doctoral studies, he was a visiting Ph.D. student at the University of Texas at Dallas, United States, in 2022, and the University of Bremen, Germany, in 2023. He has served on the organizing committees of several international conferences, including IEEE ASRU 2019, SIGDIAL 2021, and ICASSP 2022. He is also an active reviewer for multiple leading conferences and journals, such as INTERSPEECH, ICASSP, NeurIPS, SLT, ASRU, IEEE Signal Processing Letters, Speech Communication, IEEE Transactions on Affective Computing, and IEEE/ACM Transactions on Audio, Speech, and Language Processing.
\end{IEEEbiography}
\vskip-1.0em

\begin{IEEEbiography}[{\includegraphics[width=1in,height=1.25in,clip,keepaspectratio]{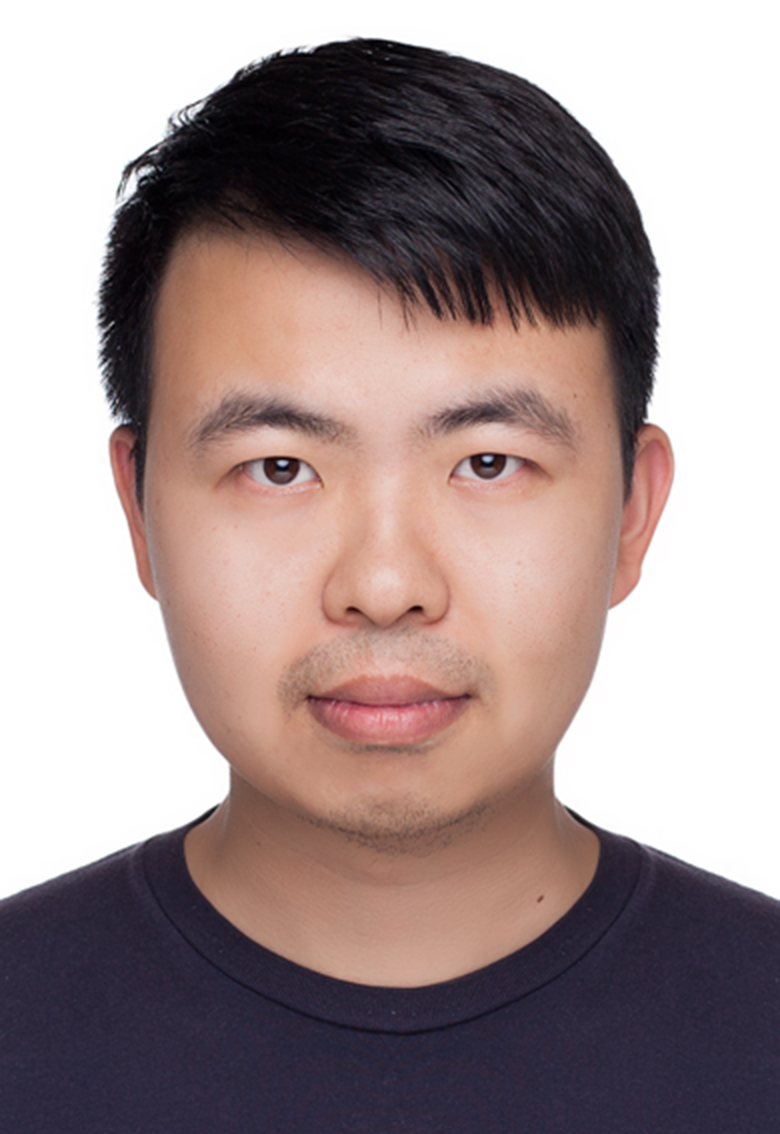}}]{Shuai Wang} (Member, IEEE) 
%obtained his Ph.D. from Shanghai Jiao Tong University (SJTU), in 2020. Currently, he serves as a research scientist at the Shenzhen Research Institute of Big Data (SRIBD). He is also an adjunct assistant professor at School of Data Science, Chinese University of Hong Kong, Shenzhen (CUHK-SZ). Prior to that, Dr. Wang worked at Tencent as a senior application scientist, leading a group working on speaker recognition, voice conversion and speech synthesis. Dr. Wang has published more than 40 papers on the topic of speaker modeling. He was the recipient of IEEE Ganesh N. Ramaswamy Memorial Student Grant Award (ICASSP2018). He was also the main contributor to the winning systems of VoxSRC 2019 and DIHARD 2019.  
%He is a member of ISCA, SPS and IEEE, searving as a regular reviewer for conferences and journals including ICASSP, INTERSPEECH, ASRU, TASLP and CSL. 
%He initiated the popular ``WeSpeaker" and ``WeSep" projects, utilized by numerous research groups across academia and industry. 
received his PhD from Shanghai Jiao Tong University (2020). He is a tenure-track associate professor at Nanjing University and holds adjunct positions at the Shenzhen Research Institute of Big Data (SRIBD) and the Chinese University of Hong Kong, Shenzhen (CUHK-SZ). He has published over 60 papers on speaker modeling and was recipient of the IEEE Ramaswamy Grant at ICASSP 2018 and winner of VoxSRC 2019 and DIHARD 2019. He is a member of ISCA, SPS and IEEE, searving as a regular reviewer for conferences and journals including ICASSP, INTERSPEECH, ASRU, TASLP and CSL. He is the initiator of the open-source projects ``WeSpeaker" and ``WeSep".
\end{IEEEbiography}

\vskip-1.0em

\begin{IEEEbiography}[{\includegraphics[width=1in,height=1.25in,clip,keepaspectratio]{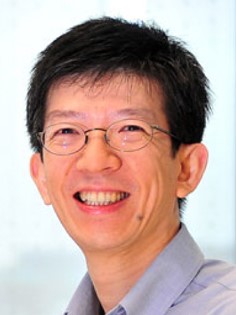}}]{
Haizhou Li % Author name
}(Fellow, IEEE) received the B.\,Sc., M.\,Sc., and Ph.D degree in electrical and electronic engineering from South China University of Technology, Guangzhou, China in 1984, 1987, and 1990 respectively. He is currently a Presidential Chair
Professor and the Executive Dean of the School of
Data Science, The Chinese University of Hong Kong,
Shenzhen, China. He is also an Adjunct Professor
with the Department of Electrical and Computer Engineering, National University of Singapore, Singapore. Prior to that, he taught with The University
of Hong Kong, Hong Kong, during 1988–1990, and South China University
of Technology, during 1990–1994. He was a Visiting Professor with CRIN,
France, during 1994–1995, Research Manager with the AppleISS Research
Centre during 1996–1998, the Research Director with Lernout \& Hauspie
Asia Pacific during 1999–2001, Vice President with InfoTalk Corporation Ltd.
during 2001–2003, and Principal Scientist and Department Head of human
language technology with the Institute for Infocomm Research, Singapore,
during 2003–2016. His research interests include automatic speech recognition,
speaker and language recognition, natural language processing. Dr. Li was
the Editor-in-Chief of IEEE/ACM TRANSACTIONS ON AUDIO , SPEECH AND
LANGUAGE PROCESSING during 2015–2018, an elected Member of IEEE Speech
and Language Processing Technical Committee during 2013–2015, the President
of the International Speech Communication Association during 2015–2017,
President of Asia Pacific Signal and Information Processing Association during
2015–2016, and President of Asian Federation of Natural Language Processing
during 2017–2018. Since 2012, he has been a Member of the Editorial Board
of Computer Speech and Language. He was the General Chair of ACL 2012,
INTERSPEECH 2014, ASRU 2019 and ICASSP 2022. Dr. Li is a Fellow of
the ISCA, and a Fellow of the Academy of Engineering Singapore. He was the
recipient of the National Infocomm Award 2002, and President’s Technology
Award 2013 in Singapore. He was named one of the two Nokia Visiting
Professors in 2009 by the Nokia Foundation, and U Bremen Excellence Chair
Professor in 2019.
\end{IEEEbiography}

\vspace{11pt}


\vfill

\end{document}